\documentclass[journal]{IEEEtran}
\usepackage{amsmath}
\usepackage{amssymb}
\usepackage{mathtools}
\usepackage{bm}
\usepackage{amsthm}
\usepackage{algorithm}
\usepackage{algpseudocode}
\usepackage{graphicx}\usepackage{tabularx}
\usepackage{array}
\usepackage{float}
\usepackage{subcaption}
\usepackage{booktabs}
\usepackage{multirow}
\usepackage{array}
\usepackage{newtxtext}
\usepackage{newtxmath}
\usepackage[dvipsnames]{xcolor}
\usepackage[colorlinks=true,
            linkcolor=blue,
            citecolor=blue,
            urlcolor=blue]{hyperref}
\usepackage{enumitem}
\usepackage[normalem]{ulem}
\usepackage{url}
\usepackage{cite}

\begin{document}

\title{Game-Theoretic Multi-Agent Control for Robust Contextual Reasoning in LLMs}

\author{
Saeid~Jamshidi,
Amin~Nikanjam$^{\ddagger}$,
Arghavan~Moradi~Dakhel,
Kawser~Wazed~Nafi and
Foutse~Khomh
\thanks{
S. Jamshidi, A. Moradi Dakhel, K. W. Nafi, and F. Khomh are with the SWAT Laboratory,
Polytechnique Montréal, Montréal, QC, Canada
(e-mail: \{saeid.jamshidi, arghavan.moradi-dakhel, kawser.wazed-nafi, foutse.khomh\}@polymtl.ca).
}%
\thanks{
A. Nikanjam$^{\ddagger}$ is with the Huawei Distributed Scheduling and Data Engine Lab,
Montréal, QC, Canada.
Work done while at Polytechnique Montréal
(e-mail: amin.nikanjam1@huawei.com).
}%

}

\maketitle

\begin{abstract}
Large Language Models (LLMs) engaged in multi-turn interactions maintain an evolving context rather than generating isolated responses, making them vulnerable to prompt-injection and context-poisoning attacks in which locally plausible adversarial fragments gradually distort reasoning trajectories. Existing defenses primarily focus on filtering individual outputs and largely ignore how context evolves across turns, leaving the dynamics of long-horizon reasoning exposed. Although the Model Context Protocol (MCP) standardizes context exchange and tool invocation, it serves as a passive routing layer and does not enforce stability in the evolution of context. To address these limitations, we introduce the Game-Theoretic Secure Model Context Protocol (GT-MCP), a controller-driven multi-agent solution that treats context management as a closed-loop dynamical process. GT-MCP coordinates three heterogeneous LLM agents and selects the outputs using a trust function that jointly evaluates causal consistency against a validated context graph, semantic agreement between agents, and distributional drift over time. When instability is detected, a rollback-based self-healing mechanism restores the validated context, preventing propagation of unsupported fragments. Empirical evaluation over 500 interaction turns under an adaptive adversarial threat model demonstrates that contextual drift remains bounded in 99.6\% of turns, with recovery required in only 0.4\%. The per-turn utility is tightly concentrated (median $=-0.19$, P05 $=-0.72$, P95 $=0.30$) with severe degradation ($<-1$) occurring only in 0.4\% of cases, and no injection attempt succeeds at the controller level. The selected outputs maintain stable win rates above 98\%, and computational overhead remains predictable (latency per token $=1.63\times10^{-3}$\,s). 
\end{abstract}


\section{Introduction}
\label{sec:intro}
Large Language Models (LLMs) are increasingly being deployed as interactive reasoning systems that operate in evolving contexts rather than on isolated prompts \cite{naveed2025comprehensive, chen2024large, li2025system}. Modern LLM-based applications combine user instructions, retrieved knowledge, tool outputs, and prior responses into persistent context streams that directly impact subsequent outputs \cite{hou2025llm, ray2025survey}. This shift enables complex multi-step reasoning, yet it also introduces a critical security weakness: adversarial fragments that appear locally plausible can persist across turns, accumulate within the context, and gradually steer the reasoning trajectory away from the intended task \cite{asimopoulosbeyond}. Recent studies show that prompt injection is not limited to explicit jailbreak attempts. Adversarial instructions can be embedded within benign-looking contextual inputs to bypass safety filters \cite{liu2024formalizing}. Prompt injection strategies are commonly categorized into direct, indirect, and multi-turn attacks \cite{geng2025survey}. In retrieval-augmented and tool-integrated systems, external documents and API responses may contain content that persists across memory updates \cite{ferrag2025indirect}. Furthermore, such attacks can remain dormant across multiple interactions before impacting downstream reasoning, thereby evading single-step detection \cite{jiang2025agentlab}.

Existing defense mechanisms address this problem mainly at the level of individual inputs and outputs. Instruction isolation, prompt sanitization, structured prompting, and post-hoc verification can reduce immediate policy violations \cite{liu2024formalizing, geng2025survey, chen2025struq}, yet they do not directly regulate how context evolves after each accepted response. This is a limitation because deployed LLM systems behave as closed-loop dynamical processes: each selected output can update the contextual state that shapes future generations \cite{hou2025model}. In this setting, prompt injection becomes a trajectory-steering problem rather than a single-response safety problem. Robustness consequently requires mechanisms that stabilize the evolution of context over time. This requirement becomes even more important in multi-model deployments, where correlated responses and compromised agents can manipulate consensus without triggering local alarms \cite{sasal2025prompt}.

This paper addresses this gap by introducing \emph{GT-MCP: Game-Theoretic Multi-Agent Control for Robust Contextual Reasoning in LLMs}. Although the name retains MCP terminology to emphasize secure context exchange, the main contribution is a trajectory-level context-control layer for multi-agent LLM systems. GT-MCP treats contextual reasoning as a disturbed dynamical process whose state must remain close to a validated reasoning manifold. This manifold is represented through a causal context graph that captures structurally grounded claims and their support relations. The controller then coordinates heterogeneous LLM agents through a trust-based selection rule that jointly evaluates causal consistency, cross-agent agreement, and distributional drift. When instability is detected, rollback-based self-healing restores the validated context and prevents low-support adversarial fragments from propagating across turns. The central idea is that robust LLM interaction requires active governance of reasoning trajectories rather than passive acceptance of context updates. Conceptually, GT-MCP integrates three components. First, context evolution is modeled as a bounded adversarial control problem. Second, multi-agent coordination is formulated as a repeated Stackelberg interaction between a controller and an adaptive adversary. Third, trust-weighted selection combines structural grounding, semantic agreement, and temporal stability to make adversarial deviation strategically unattractive. The main contributions of this work are as follows:
\begin{itemize}
    \item We formulate secure contextual reasoning in LLM-based systems as a \emph{controlled dynamical process} and introduce GT-MCP, a trajectory-level context-control solution that stabilizes reasoning through structural validation, distributional drift monitoring, and rollback-based recovery.
    
    \item We design a \emph{game-theoretic mechanism} in which a controller coordinates heterogeneous LLM agents using a trust-based selection rule that jointly incorporates causal grounding, cross-agent agreement, and temporal stability, shaping incentive-compatible behavior under adversarial prompt injection.
    
    \item We evaluate GT-MCP across 500 interaction turns under an adaptive adversarial threat model, showing bounded contextual drift in 99.6\% of turns, rare recovery activation in 0.4\% of turns, zero controller-level injection success, stable selected-output win rates above 98\%, and predictable efficiency scaling.
\end{itemize}

The remainder of this paper is organized as follows. Section~\ref{sec:related_work} reviews related work. Section~\ref{sec:methodology} presents GT-MCP, including the threat model, causal context graph, trust-based coordination, and recovery mechanisms. Section~\ref{sec:results} reports experimental results on stability, utility, efficiency, and robustness. Section~\ref{sec:discussion} discusses the findings, Section~\ref{sec:limitations_future} outlines limitations and future directions, and Section~\ref{Conclusion} concludes the paper.

\section{Related Work}
\label{sec:related_work}
 This section reviews prior work along four directions and identifies the remaining gap in trajectory-level context control.

\subsection{Prompt Injection and Context Poisoning}
Prompt injection has evolved from single-turn jailbreaks toward \emph{stateful} failures in which injected instructions persist inside long-lived contexts, retrieved passages, and tool outputs. Early optimization-based jailbreak work shows that short adversarial strings can reliably disrupt aligned behavior without modifying model weights \cite{zou2023gcg}. More recent studies focus on \emph{indirect} prompt injection, in which attackers embed instructions in external content that later enter the model's context via retrieval and tool channels. Liu \emph{et al.} \cite{liu2023bipia} formalize indirect injection settings and evaluate defenses under realistic constraints, showing that prompt-boundary filtering alone is insufficient when malicious content can persist across interactions. These studies establish prompt injection as a context-contamination problem, but they do not provide a control mechanism for regulating how accepted content affects future reasoning states.

\subsection{Retrieval-Augmented Generation (RAG) Corruption Attacks}
Retrieval-augmented generation introduces a distinct attack surface because the model may reason fluently over poisoned evidence. PoisonedRAG \cite{shi2024poisonedrag} demonstrates that targeted corruption of retrieved content can steer generation despite benign user prompts. This line of work highlights the need to validate provenance and structural support before incorporating retrieved claims into memory. However, most RAG-oriented defenses focus on retrieval-time filtering and evidence sanitization, whereas the longer-term evolution of the validated context following each accepted response is less directly controlled.

\subsection{Agentic Systems and Tool-Mediated Injection}
AI agents with tool integration amplify the risk of injection because tool outputs can affect subsequent reasoning, memory updates, and action selection. AgentDojo \cite{debenedetti2024agentdojo} provides a benchmark for evaluating indirect prompt injection in agentic settings, enabling controlled analysis under repeated interactions. Supporting work shows that attacks can target policies governing tool use and memory updates, motivating explicit provenance boundaries across context segments \cite{liu2024targetingcore}. These studies clarify the attack surface of agentic systems, yet they primarily assess vulnerabilities and local safeguards rather than modeling the evolution of multi-agent contexts as a controlled state-transition process.

\subsection{Defenses: Isolation, Hierarchy, and Design-Time Control}
Several defenses separate trusted instructions from untrusted content using instruction hierarchy, isolation, and design-time separation principles. Spotlighting marks untrusted segments and constrains the model's attention to them, reducing indirect injection transfer into core instructions \cite{wei2024spotlighting}. Design-time approaches advocate separating the instruction and data channels to reduce the attack surface \cite{williams2025defeating}. Related instruction-isolation methods emphasize explicit boundary enforcement and controlled merging between trusted state and untrusted inflow \cite{schutera2025instructionisolation}. System-level design patterns further provide practical guidance for preserving provenance, limiting authority escalation, and auditing state updates in LLM applications \cite{beurerkellner2025designpatterns}. These approaches are valuable for reducing immediate instruction override, but they remain primarily local: they protect boundaries and validate segments without explicitly controlling the trajectory of context over multiple turns.\\

Despite substantial progress, existing defenses do not fully address the long-term dynamics of contextual LLMs. Most prior methods focus on filtering out untrusted segments, separating the instruction and data channels, restricting tools' authority, and hardening retrieval pipelines. These mechanisms reduce the immediate attack surface, yet they do not model context as a dynamical state that can be gradually steered by adversarial perturbations once responses are accepted and merged into memory. They also do not analyze multi-agent selection under an incentive-aware adversarial setting. In contrast, our work treats secure contextual reasoning as a controlled dynamical process and introduces a game-theoretic multi-agent mechanism that jointly enforces structural grounding, temporal stability, and incentive-compatible selection. This shift from local filtering toward trajectory-level context control is the central distinction between GT-MCP and prior defenses.

\section{Methodology: Game-Theoretic Secure Context Control (GT-MCP)}
\label{sec:methodology}
We propose \emph{Game-Theoretic Secure Context Control (GT-MCP)}, a trajectory-level control layer designed to stabilize multi-agent LLM reasoning under adversarial contextual perturbations. GT-MCP models multi-turn interaction as a closed-loop context-control process in which accepted outputs influence subsequent reasoning states. At interaction turn $t \in \{1,\dots,T\}$, the user query is denoted by $q_t$, and the controller maintains a validated context state $c_t$. This state contains trusted semantic memory, including verified claims, validated summaries, retrieved evidence, and previously accepted reasoning elements. The observed context $\tilde{c}_t$ provided to the agents may additionally include untrusted inflows $\Delta c_t$, such as injected prompts, poisoned retrieval passages, manipulated tool outputs, and externally supplied contextual fragments:
\begin{equation}
\tilde{c}_t = c_t \oplus \Delta c_t.
\end{equation}
Here, $\oplus$ denotes structured concatenation with provenance markers. This representation preserves the separation between validated memory and untrusted inflows, enabling later auditing, quarantine, and selective removal of unsupported fragments.
\begin{figure*}[t]
\centering
\includegraphics[width=0.8\textwidth]{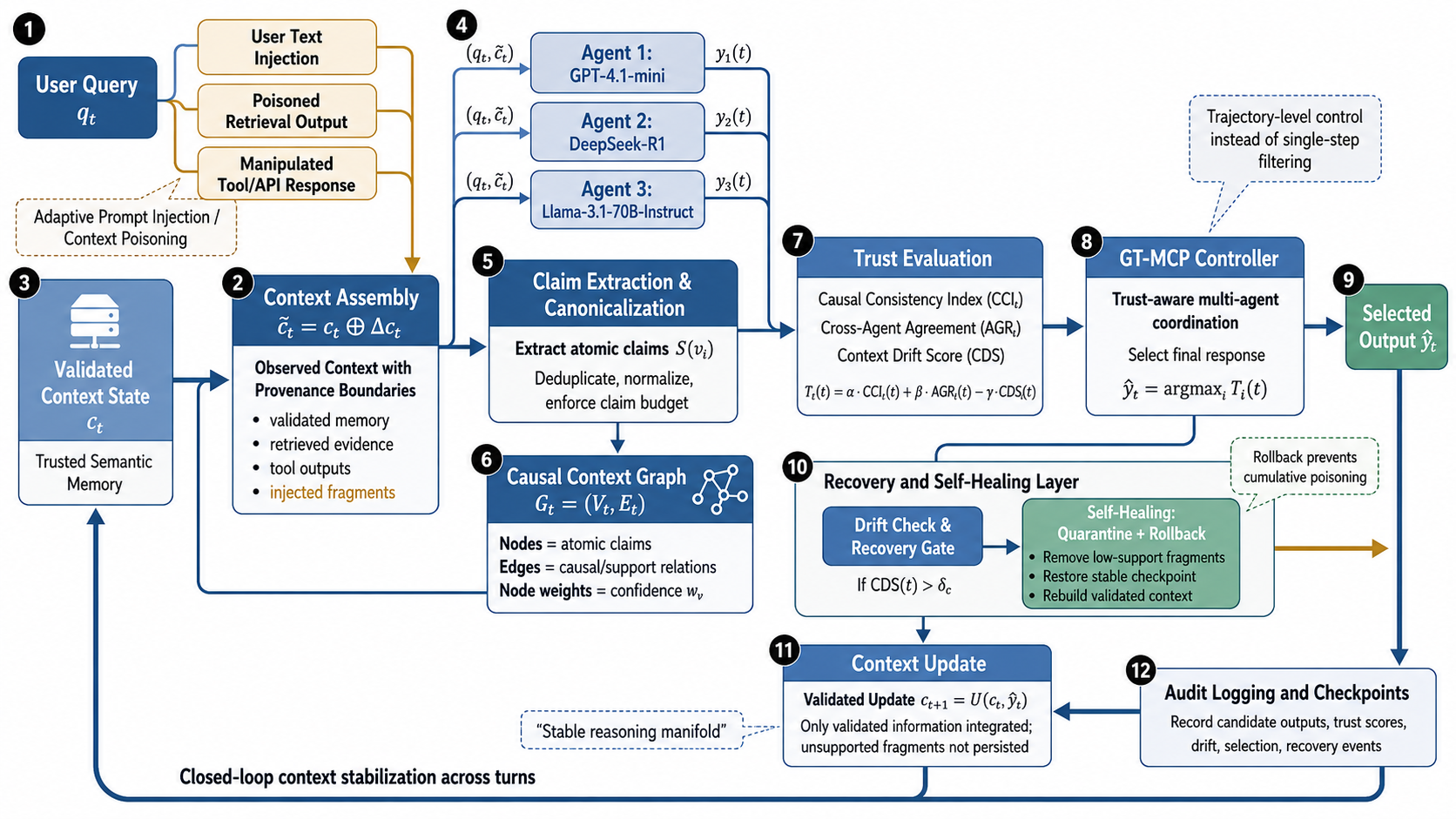}
\caption{Layered GT-MCP architecture for closed-loop context stabilization. The controller separates validated context from untrusted inflows, coordinates heterogeneous LLM agents, scores candidate outputs based on causal consistency, cross-agent agreement, and contextual drift, and updates persistent memory only through validated selections.}
\label{fig:gtmcp_architecture}
\end{figure*}
As illustrated in Figure~\ref{fig:gtmcp_architecture}, GT-MCP first constructs the observed context $\tilde{c}_t$ from the validated state $c_t$, the query $q_t$, retrieved evidence, and possible adversarial perturbations. The same input pair $(q_t,\tilde{c}_t)$ is then provided to three heterogeneous LLM agents, each producing a candidate response $y_i(t)$. The controller extracts atomic claims from each candidate, canonicalizes them, maps them into a causal context graph $G_t=(V_t,E_t)$, and assigns trust scores using three complementary signals: causal consistency, cross-agent agreement, and contextual drift. The response with the highest trust score is selected as $\hat{y}_t$ and is used to update the validated context through the legitimate update operator:
\begin{equation}
c_{t+1}=\mathcal{U}(c_t,\hat{y}_t).
\end{equation}
This update rule ensures that persistent memory evolves only through controller-approved outputs. If the contextual drift score exceeds the recovery threshold, the self-healing layer quarantines low-support fragments, restores a stable checkpoint, and rebuilds the validated context before the next turn. The controller also records candidate outputs, extracted claims, trust scores, drift values, selected responses, and recovery events in audit logs. Through this closed-loop design, GT-MCP prevents unsupported contextual fragments from silently propagating across turns and treats context evolution as a security-critical dynamical process.

\subsection{Notation and Preliminaries}
\label{subsec:notation}
GT-MCP is formulated as a multi-agent decision process in which three heterogeneous LLMs reason over a shared, evolving context. At turn $t \in \{1,\dots,T\}$, the user query is denoted by $q_t$, and the controller maintains a validated context state $c_t$. This state contains trusted semantic memory, including verified claims, validated summaries, retrieved evidence, and previously accepted reasoning elements. The observed context $\tilde{c}_t$ provided to the agents may additionally contain untrusted contextual perturbations $\Delta c_t$:
\[
\tilde{c}_t = c_t \oplus \Delta c_t .
\]
Each agent $i \in \{1,2,3\}$ generates a candidate response conditioned on the query and observed context:
\begin{equation}
y_i(t)\sim p_i\big(y \mid q_t,\tilde{c}_t\big),
\end{equation}
where $p_i(\cdot)$ denotes the response distribution induced by agent $i$ under contextual uncertainty and possible adversarial influence. The controller selects a final output $\hat{y}_t$ from the candidate set $\{y_1(t),y_2(t),y_3(t)\}$ and updates the validated state only through the legitimate update operator $\mathcal{U}$:
\[
c_{t+1}=\mathcal{U}(c_t,\hat{y}_t).
\]
To evaluate whether candidate outputs are structurally grounded, the controller maintains a causal context graph $G_t=(V_t,E_t)$, where each node $v\in V_t$ represents an atomic claim and each edge $(u\rightarrow v)\in E_t$ encodes entailment, causal support, and dependency between claims. Each node is assigned a confidence weight $w_v\in[0,1]$. For a candidate response $y_i(t)$, the extracted claim set is denoted by $\mathcal{S}(y_i(t))$. The controller computes three primary signals: the causal consistency index $\mathrm{CCI}_i(t)$, which measures structural support from $G_t$; the agreement score $\mathrm{AGR}_i(t)$, which measures semantic consistency with peer agents through the kernel $\kappa(\cdot,\cdot)$; and the contextual drift score $\mathrm{CDS}(t)$, which measures semantic deviation between $\tilde{c}_t$ and $c_t$. Unless stated otherwise, expectations are taken into account for model stochasticity, adversarial randomness, and uncertainty in claim extraction.
\begin{table}[t]
\centering
\caption{Key notation used in the GT-MCP methodology.}
\label{tab:notation}
\begin{tabular}{ll}
\toprule
Symbol & Meaning \\
\midrule
$t, T$ & Interaction turn and horizon length \\
$q_t$ & User query at turn $t$ \\
$c_t$ & Validated context state \\
$\tilde{c}_t$ & Observed context provided to agents \\
$\Delta c_t$ & Untrusted contextual perturbation \\
$\oplus$ & Structured concatenation with provenance markers \\
$y_i(t)$ & Candidate response generated by agent $i$ \\
$\hat{y}_t$ & Controller-selected output \\
$\mathcal{U}(\cdot)$ & Legitimate context update operator \\
$G_t=(V_t,E_t)$ & Causal context graph \\
$v, w_v$ & Atomic claim node and confidence weight \\
$\mathcal{S}(y)$ & Extracted claim set from response $y$ \\
$\mathrm{CCI}_i(t)$ & Causal consistency index of agent $i$ \\
$\mathrm{AGR}_i(t)$ & Cross-agent agreement score of agent $i$ \\
$\mathrm{CDS}(t)$ & Contextual drift score \\
$\kappa(\cdot,\cdot)$ & Semantic agreement kernel \\
$T_i(t)$ & Trust score assigned to candidate $i$ \\
$\alpha,\beta,\gamma$ & Trust aggregation weights \\
$\ell(\hat{y}_t)$ & Output risk function \\
$\psi(\Delta c_t)$ & Semantic anomaly measure \\
$\delta_c$ & Recovery threshold for self-healing \\
$\pi$ & Controller selection policy \\
$\mathcal{Y}_{adv}$ & Set of adversarially successful outputs \\
\bottomrule
\end{tabular}
\end{table}

\subsection{Design Rationale}
\label{subsec:rationale}
GT-MCP is motivated by the observation that prompt injection in multi-turn LLM systems is not only a local input-output failure but also a trajectory-level state-transition problem. In persistent-context applications, a response accepted at turn $t$ may become part of the memory that conditions later generations. Therefore, an adversarial fragment that appears locally plausible can gradually shift the reasoning trajectory if it is repeatedly accepted, summarized, and used as support for future claims. GT-MCP addresses this risk by treating multi-agent LLM interaction as a controlled reasoning process in which the context state must remain structurally grounded and temporally stable under adversarial perturbations. The controller evaluates candidate responses using three complementary signals. First, \emph{structural grounding} measures whether extracted claims are supported by the validated causal context graph $G_t$. Second, \emph{cross-agent agreement} assesses whether a candidate is semantically consistent with other agents, thereby reducing the influence of unreliable. Third, \emph{temporal stability} measures whether incoming contextual information induces semantic drift away from the validated state. In contrast to simple voting, this design does not assume that agreement alone implies correctness; a candidate must also be grounded in the validated graph and must not introduce excessive contextual drift. Let $\{c_t^\star\}_{t=1}^{T}$ denote an ideal validated context trajectory and $\{c_t\}_{t=1}^{T}$ denote the trajectory maintained by the controller. Long-horizon contextual stability is measured by the cumulative deviation:
\begin{equation}
\mathcal{J}_{\mathrm{stab}}
=
\mathbb{E}
\left[
\sum_{t=1}^{T}
d(c_t,c_t^\star)
\right],
\end{equation}
where $d(\cdot,\cdot)$ denotes a semantic distance between contextual states, implemented through embedding-based distances and distributional distances over extracted claim sets. This objective captures how far the maintained context drifts from the desired validated trajectory over multiple turns.
To jointly account for local response safety and long-horizon stability, the controller minimizes:
\begin{equation}
\mathcal{J}
=
\mathbb{E}
\left[
\sum_{t=1}^{T}
\left(
\ell(\hat{y}_t)
+
\mu\, d(c_t,c_t^\star)
\right)
\right],
\end{equation}
where $\ell(\hat{y}_t)$ penalizes unsafe, unsupported, and adversarially influenced selected outputs, and $\mu>0$ controls the trade-off between immediate output risk and trajectory-level stability. This formulation reflects the central design principle of GT-MCP: a response should not be selected because it is fluent; it should be selected only when it is structurally supported, consistent with peer agents, and compatible with the validated context trajectory. At each turn, the three agents receive the same observed input pair $(q_t,\tilde{c}_t)$ and generate candidate responses:
\begin{equation}
y_i(t)\sim p_i(y\mid q_t,\tilde{c}_t),
\quad i\in\{1,2,3\}.
\end{equation}
The controller then scores the candidates using causal consistency, cross-agent agreement, and contextual drift before selecting $\hat{y}_t$. Only the selected and validated output is allowed to update the persistent context. This closed-loop design reduces the likelihood that unsupported adversarial fragments accumulate across turns and provides the basis for trajectory-level robustness in GT-MCP.

\subsection{Context Dynamics}
\label{subsec:dynamics}
At each interaction turn, the system receives information that may contain adversarial perturbations $\Delta c_t$ originating from user prompts, retrieved passages, and tool outputs. The observed context provided to the agents is:
\begin{equation}
\tilde{c}_t = c_t \oplus \Delta c_t,
\end{equation}
where $c_t$ denotes the validated context maintained by the controller and $\Delta c_t$ represents untrusted inflows. This decomposition preserves provenance boundaries between trusted semantic memory and externally introduced content, allowing suspicious fragments to be audited, quarantined, and removed without directly modifying the validated state. After the controller selects the validated output $\hat{y}_t$, the trusted context evolves according to:
\begin{equation}
c_{t+1} = \mathcal{U}(c_t,\hat{y}_t),
\end{equation}
where $\mathcal{U}(\cdot)$ denotes the legitimate update operator, including verified-claim insertion, summarization, checkpoint updates, and graph refinement. This equation ensures that only validated outputs may modify persistent memory. Incoming perturbations at the next turn remain confined to the observed context until they satisfy structural and temporal validation constraints. If abnormal drift and low-support structures are detected, the recovery layer triggers rollback and quarantine procedures to restore a stable context state. To model realistic prompt injection, perturbations are constrained to remain locally plausible rather than trivially detectable:
\begin{equation}
\psi(\Delta c_t)\le\epsilon,
\end{equation}
where $\psi(\cdot)$ measures semantic anomaly using signals such as topic deviation, contradiction density, and embedding outlier scores. Under this constraint, adversarial fragments may appear semantically coherent at the local level, thereby forcing the controller to rely on structural grounding, cross-agent agreement, and temporal consistency rather than superficial lexical filtering. Consequently, GT-MCP treats prompt injection as a long-horizon context-stabilization problem rather than a single-step classification problem.

\subsection{Adversarial Game Formulation}
\label{subsec:adversarial_game}
GT-MCP formulates prompt injection and context poisoning as a repeated interaction between a controller and an adaptive attacker. The attacker may influence the observed context through malicious user prompts, poisoned retrieved content, manipulated tool outputs, and compromised candidate generations. However, the attacker cannot directly overwrite the validated context state $c_t$; it can only introduce perturbations $\Delta c_t$ into the observed context $\tilde{c}_t$. The attacker succeeds only if these perturbations cause the controller to select an adversarial output. Let $\mathcal{Y}_{adv}$ denote the set of adversarially successful outputs. This set includes outputs that bypass controller constraints, introduce unsupported claims into the validated context, manipulate tool behavior, leak protected information, and gradually steer the maintained context away from the ideal validated trajectory. The attacker aims to maximize the number of controller-level failures over the interaction horizon:
\begin{equation}
\max_{\{\Delta c_t\}_{t=1}^{T}}
\sum_{t=1}^{T}
\mathbb{I}
\left(
\hat{y}_t \in \mathcal{Y}_{adv}
\right),
\end{equation}
where $\mathbb{I}(\cdot)$ is the indicator function and $\hat{y}_t$ is the controller-selected output at turn $t$. The controller seeks a selection-and-recovery policy $\pi$ that remains robust under adaptive, history-dependent perturbations. This yields the following minimax objective:
\begin{equation}
\min_{\pi}
\max_{\{\Delta c_t\}_{t=1}^{T}}
\mathcal{J}\big(\pi,\{\Delta c_t\}_{t=1}^{T}\big),
\end{equation}
where $\pi$ maps the observed context $\tilde{c}_t$ and candidate responses $\{y_i(t)\}_{i=1}^{3}$ to the selected output $\hat{y}_t$. The objective $\mathcal{J}$ combines local output risk and long-horizon contextual deviation, as defined in Section~\ref{subsec:rationale}. This interaction can be interpreted as a repeated Stackelberg security game \cite{zychowski2026scalability}, in which the controller commits to a trust-and-recovery policy, and the attacker responds by generating plausible perturbations over multiple turns. This framing is appropriate because adversarial fragments may remain locally coherent while gradually shifting the contextual trajectory. Therefore, robustness must be evaluated not only as single-turn attack detection but also as long-horizon control of context evolution.

\subsection{Threat Model Details}
\label{subsec:threat}
GT-MCP assumes an adaptive adversary that cannot directly overwrite the validated context state $c_t$, but can influence the observed context through untrusted perturbations:
\begin{equation}
\tilde{c}_t = c_t \oplus \Delta c_t,
\end{equation}
where $\Delta c_t$ may include malicious user instructions, poisoned retrieved passages, manipulated tool outputs, hidden directives embedded in external content, and adversarial candidate generations. Only controller-approved outputs $\hat{y}_t$ can update the trusted state through the legitimate update operator $\mathcal{U}(\cdot)$. Thus, the adversarial impact is indirect: the attacker must cause a malicious, unsupported candidate to be selected before it can affect persistent memory.
Table~\ref{tab:attack_taxonomy} summarizes the attack families considered in this work. These attacks differ in their injection channels, temporal behavior, and visibility, but they share the same objective: to influence future reasoning by inserting unsupported and adversarial fragments into the trajectory of controller-selected outputs.
\begin{table*}[t]
\centering
\caption{Attack taxonomy considered in GT-MCP. Each attack enters through the untrusted perturbation term $\Delta c_t$ and attempts to affect future reasoning through controller-selected outputs.}
\label{tab:attack_taxonomy}
\begin{tabular}{p{2.8cm} p{2.6cm} p{3.0cm} p{5.2cm} p{2.6cm}}
\toprule
Family & Injection channel & Temporal pattern & Representative objective & Visibility \\
\midrule
Direct injection & User text ($q_t$) & Immediate & Override controller constraints; induce policy bypass; force tool misuse & Explicit \\
Retrieval poisoning & Retrieved passages ($r_t$) & Immediate / delayed & Insert hidden instructions into evidence; corrupt reasoning through poisoned support & Often implicit \\
Tool-output injection & Tool/API outputs & Immediate & Escalate tool authority; inject high-trust directives; manipulate memory updates & Implicit \\
Dormant trigger & Any channel & Multi-turn latent & Plant benign-looking content that activates after $k$ turns & Implicit \\
Trajectory steering & Any channel & Gradual & Slowly shift task framing, constraints, and reasoning state & Implicit \\
Agreement mimicry & Byzantine agent & Multi-turn & Mimic peer agreement while inserting unsupported claims & Implicit \\
\bottomrule
\end{tabular}
\end{table*}
Adversarial perturbations are sequential and history-dependent:
\begin{equation}
\Delta c_t = \mathcal{A}_t(\mathcal{H}_t),
\qquad
\mathcal{H}_t = \{(q_s,\hat{y}_s)\}_{s=1}^{t-1},
\end{equation}
where $\mathcal{A}_t$ denotes the attacker strategy at turn $t$, and $\mathcal{H}_t$ is the observable interaction history available to the attacker. Although $\Delta c_t$ cannot directly modify $c_t$, it may indirectly affect future reasoning if it causes the controller to select an adversarial output:
\begin{equation}
c_{t+1} = \mathcal{U}(c_t,\hat{y}_t).
\end{equation}
We consider both \emph{myopic} and \emph{strategic} attackers. A myopic attacker attempts to maximize immediate controller-level failure while remaining locally plausible under a bounded anomaly constraint:
\begin{equation}
\psi(\Delta c_t)\le\epsilon,
\end{equation}
where $\psi(\cdot)$ measures semantic anomaly using signals such as topic deviation, contradiction density, and embedding-level outlier behavior. A strategic attacker optimizes delayed impact over multiple turns:
\begin{equation}
\max_{\{\Delta c_\tau\}_{\tau=t}^{T}}
\sum_{\tau=t}^{T}
\rho_d^{\tau-t}
\mathbb{I}
\left(
\hat{y}_\tau \in \mathcal{Y}_{adv}
\right),
\qquad
\psi(\Delta c_\tau)\le\epsilon,
\end{equation}
where $\rho_d\in(0,1]$ is a temporal discount factor. This setting captures latent triggers and gradual contextual steering that may appear benign at any individual turn but become harmful over a longer horizon. To model compromised generation behavior, we additionally assume that at most one of the three LLM agents may behave adversarially at a given turn:
\begin{equation}
y_{i^\dagger}(t)\sim \mathcal{A}^{LLM}_t,
\qquad
y_i(t)\sim p_i(y\mid q_t,\tilde{c}_t),
\quad
i\neq i^\dagger.
\end{equation}
This one-Byzantine-agent assumption prevents the defense from relying on majority voting alone and motivates the need for structural grounding and drift monitoring. Controller-level adversarial success is defined as:
\begin{equation}
\mathbb{I}_{adv}(t)=
\mathbb{I}
\left(
\hat{y}_t\in\mathcal{Y}_{adv}
\right).
\end{equation}
A selected output belongs to $\mathcal{Y}_{adv}$ if it bypasses controller constraints, injects unsupported claims into the validated graph, manipulates tool behavior, leaks protected information, and causes persistent contextual drift without triggering recovery. The long-horizon attacker's objective is therefore:
\begin{equation}
\max_{\{\Delta c_t\}_{t=1}^{T}}
\mathbb{E}
\left[
\sum_{t=1}^{T}
\mathbb{I}_{adv}(t)
\right].
\end{equation}
This threat model captures both immediate injection failures and delayed trajectory manipulation, aligning the evaluation of GT-MCP with the long-horizon stability of persistent-context LLM systems.

\subsection{Game-Theoretic Formulation of Secure Multi-Agent Interaction}
\label{subsec:stackelberg}
GT-MCP models secure multi-agent reasoning as a repeated Stackelberg-style interaction between a controller and an adaptive attacker. At each turn $t$, the controller commits to a selection-and-recovery policy $\pi$ that maps the observed context $\tilde{c}_t$ and the candidate responses $\{y_i(t)\}_{i=1}^{3}$ to a selected output $\hat{y}_t$. The attacker observes the interaction history and attempts to steer future context states by introducing bounded perturbations, $\Delta c_t$, to the observed context. The interaction is therefore strategic: the controller aims to preserve contextual stability, while the attacker attempts to induce selected outputs that corrupt the validated trajectory.
The defender utility at turn $t$ is defined as:
\begin{equation}
U_D(t)
=
-\ell(\hat{y}_t)
-
\mu\, d(c_t,c_t^\star),
\end{equation}
where $\ell(\hat{y}_t)$ penalizes unsafe, unsupported, and adversarially influenced selected outputs, and $d(c_t,c_t^\star)$ measures deviation from the ideal validated trajectory. The parameter $\mu>0$ controls the trade-off between local output risk and long-horizon contextual stability. The attacker utility is defined as:
\begin{equation}
U_A(t)
=
\mathbb{I}
\left(
\hat{y}_t\in\mathcal{Y}_{adv}
\right)
-
\rho\,\psi(\Delta c_t),
\end{equation}
where the first term rewards controller-level adversarial success and the second term penalizes perturbations that are too anomalous to remain plausible. Here, $\rho>0$ controls the cost of semantic anomaly. The corresponding discounted returns are:
\begin{align}
V_D &= \sum_{t=1}^{T}\rho_D^{t-1}U_D(t), \\
V_A &= \sum_{t=1}^{T}\rho_A^{t-1}U_A(t),
\end{align}
where $\rho_D,\rho_A\in(0,1]$ are defender and attacker discount factors. These returns reflect that both the defender and the attacker optimize over trajectories rather than isolated responses.
The controller selects the output with the highest trust score:
\begin{equation}
\hat{y}_t
=
y_{\hat{i}}(t),
\qquad
\hat{i}
=
\arg\max_{i\in\{1,2,3\}} T_i(t),
\end{equation}
where $T_i(t)$ combines causal consistency, cross-agent agreement, and contextual drift penalties. This selection rule creates an incentive structure in which adversarial candidates must satisfy multiple constraints simultaneously: they must appear structurally supported by the causal context graph, remain semantically consistent with peer agents, and avoid introducing excessive contextual drift. A candidate that is fluent but weakly grounded receives a low causal consistency score; a candidate that mimics agreement without structural support remains vulnerable to graph-based validation; and a candidate that gradually shifts the trajectory is penalized through contextual drift. Under the one-Byzantine-agent assumption and bounded perturbation constraint $\psi(\Delta c_t)\le\epsilon$, agreement alone is insufficient for adversarial success. The compromised agent cannot achieve pairwise agreement on its own, and unsupported claims cannot attain high trust unless they are also structurally grounded in $G_t$. Consequently, the controller reduces the expected payoff of adversarial deviation by assigning lower trust to candidates that lack causal support, contradict high-confidence graph nodes, and induce abnormal drift. In operational terms, this means that adversarial success requires simultaneous evasion of the causal consistency, agreement, and drift-monitoring components. This formulation establishes the GT-MCP's incentive-alignment principle. Structurally grounded and agreement-consistent candidates are rewarded with higher selection probability, whereas adversarial candidates incur structural, semantic, and temporal penalties. The controller, therefore, acts not only as an output selector but also as a strategic regulator of long-horizon context evolution.

\subsection{Causal Context Graph}
\label{subsec:graph}
To enforce structural grounding, GT-MCP maintains a directed \emph{causal context graph}:
\begin{equation}
G_t=(V_t,E_t),
\end{equation}
where each node $v\in V_t$ represents an atomic claim extracted from the validated context and from candidate outputs, and each directed edge $(u\rightarrow v)\in E_t$ represents entailment, dependency, and causal support between claims. The graph provides a structured memory of validated reasoning elements and allows the controller to distinguish supported claims from plausible but unsupported adversarial fragments. Each node is assigned a confidence weight $w_v\in[0,1]$:
\begin{equation}
w_v=\sigma\big(\mathrm{support}(v)\big),
\end{equation}
where $\sigma(\cdot)$ maps aggregated support evidence into a bounded confidence score. The support of claim $v$ is computed as:
\begin{equation}
\begin{aligned}
\mathrm{support}(v)
&=
\omega_1\,\mathbf{1}[\mathrm{persist}(v)]
+
\omega_2
\frac{1}{|\mathrm{Pa}(v)|+\epsilon}
\sum_{u\in \mathrm{Pa}(v)} w_u
\\
&\quad
+
\omega_3\,\mathrm{agree}(v)
-
\omega_4\,\mathrm{contradict}(v).
\end{aligned}
\end{equation}
where $\omega_1,\omega_2,\omega_3,\omega_4 \ge 0$ are weighting coefficients, $\epsilon>0$ prevents division by zero, and $\mathrm{Pa}(v)$ denotes the parent claims supporting $v$. The persistence term $\mathbf{1}[\mathrm{persist}(v)]$ indicates repeated appearance across validated checkpoints. The normalized parent-support term measures how strongly $v$ is supported by previously validated claims. The agreement term $\mathrm{agree}(v)$ captures claim-level consistency across agents, whereas $\mathrm{contradict}(v)$ penalizes conflicts with high-confidence nodes. Under the one-Byzantine-agent assumption, agreement is counted only when supported by at least two structurally grounded agents, preventing a single compromised model from inflating the confidence of an unsupported claim.
Atomic claims are extracted using a controller-side operator:
\begin{equation}
\mathcal{S}(y)=\mathcal{E}(y),
\end{equation}
where $\mathcal{E}(\cdot)$ maps a candidate response into a set of canonical claims. Each claim is represented as a normalized subject--predicate--object proposition and assigned a modality tag, such as fact, hypothesis, recommendation, and constraint. GT-MCP enforces a fixed claim budget:
\begin{equation}
|\mathcal{S}(y)|\le B,
\end{equation}
which prevents verbosity-based manipulation, repeated paraphrasing, and artificial inflation of graph support. Claims are canonicalized through entity normalization, lemmatization, lowercasing, deduplication, and hedging normalization, producing stable identifiers $\mathrm{id}(v)$. Non-groundable claims, including meta-instructions, self-referential directives, unverifiable tool claims, and unsupported policy-changing statements, are tagged and assigned a low default support level. This prevents instruction-like fragments from entering the validated graph merely because they are fluent and repeated. Edges are constructed in two stages. First, the controller retrieves candidate parent nodes using embedding similarity and shared entities, selecting the top-$k$ potential parents for each extracted claim. Second, a natural language inference module $\mathcal{N}(\cdot,\cdot)$ evaluates entailment, neutrality, and contradiction between the candidate claim and its potential parents. An edge is inserted only when entailment exceeds a predefined threshold, and no high-confidence contradiction is detected. Explicit causal markers, such as ``because'', ``therefore'', and mechanism-oriented phrasing, are used as auxiliary signals for edge directionality. Only claims with confidence weight $w_v\ge \tau_w$ are eligible for insertion into the next validated graph $G_{t+1}$. Claims linked to quarantined fragments retain provenance pointers, enabling fine-grained rollback of affected subgraphs during self-healing. In this way, the causal context graph acts as the structural memory of GT-MCP: it supports validated reasoning, suppresses unsupported adversarial fragments, and provides the controller with an auditable basis for trust scoring and recovery.

\subsection{CCI}
\label{subsec:cci}
For each candidate response $y_i(t)$, let $\mathcal{S}(y_i(t))$ denote the set of extracted atomic claims. The CCI measures the extent to which these claims are structurally supported by the validated causal context graph $G_t$. It is defined as:
\begin{equation}
\mathrm{CCI}_i(t)
=
\frac{1}{|\mathcal{S}(y_i(t))|+\epsilon}
\sum_{v \in \mathcal{S}(y_i(t))}
\phi(v,G_t),
\end{equation}
where $\epsilon>0$ prevents numerical instability when no extractable claim is available. The claim-level support score $\phi(v,G_t)$ is computed as:
\begin{equation}
\phi(v,G_t)
=
\min
\left(
1,\,
\frac{1}{|\mathrm{Pa}(v)|+\epsilon}
\sum_{u\in \mathrm{Pa}(v)} w_u
\right),
\end{equation}
where $\mathrm{Pa}(v)$ denotes the parent set of claim $v$ in $G_t$, and $w_u$ is the confidence weight of parent claim $u$. The CCI therefore lies in the interval $[0,1]$ and assigns higher scores to claims supported by high-confidence, validated parents. Unsupported, isolated, and contradictory claims receive low support and reduce the candidate's overall causal consistency. This metric is especially useful for detecting adversarial fragments that are fluent and locally plausible but lack structural support within the validated reasoning trajectory. Because CCI is computed after claim extraction and graph matching, it assesses whether a candidate can be integrated into the existing validated context rather than merely whether it appears semantically coherent.

\subsection{Distributional Stability Metric}
\label{subsec:cds}
Although CCI evaluates local structural support at the claim level, GT-MCP also monitors global semantic stability across interaction turns. Let $P_t$ denote the semantic distribution of the validated context $c_t$, and let $\tilde{P}_t$ denote the semantic distribution induced by the observed context $\tilde{c}_t$. These distributions are estimated over embedding clusters and topic-level claim representations. The CDS measures the distributional deviation between the observed and validated contexts. To obtain a bounded and symmetric stability measure, CDS is defined using Jensen--Shannon divergence:
\begin{equation}
\mathrm{CDS}(t)
=
D_{\mathrm{JS}}
\left(
\tilde{P}_t \parallel P_t
\right),
\end{equation}
where
\begin{equation}
D_{\mathrm{JS}}
\left(
\tilde{P}_t \parallel P_t
\right)
=
\frac{1}{2}
D_{\mathrm{KL}}
\left(
\tilde{P}_t \parallel M_t
\right)
+
\frac{1}{2}
D_{\mathrm{KL}}
\left(
P_t \parallel M_t
\right),
\end{equation}
and
\begin{equation}
M_t=\frac{1}{2}\left(\tilde{P}_t+P_t\right).
\end{equation}
Small smoothing constants are applied to the empirical distributions before normalization to avoid zero-probability bins. A low CDS indicates that the observed context remains semantically aligned with the validated state, whereas a high CDS indicates that incoming information deviates from the established reasoning trajectory. In GT-MCP, CDS complements CCI: CCI assesses whether individual claims are structurally grounded in the causal context graph, whereas CDS assesses whether the overall context distribution remains stable across turns. This combination enables the controller to detect both abrupt contradiction-based perturbations and gradual trajectory-steering attacks that may appear locally coherent but become destabilizing over multiple interactions.

\subsection{Strategic Trust Aggregation}
\label{subsec:trust}
After computing structural grounding and semantic stability signals, the controller assigns each candidate response a trust score. In contrast to simple majority voting, the trust score evaluates whether a candidate can be safely integrated into the validated context. For each candidate $y_i(t)$, the controller first constructs a tentative context update:
\begin{equation}
c_{t+1}^{(i)}=\mathcal{U}(c_t,y_i(t)),
\end{equation}
and computes a candidate-specific drift score:
\begin{equation}
\mathrm{CDS}_i(t)
=
D_{\mathrm{JS}}
\left(
P_{t+1}^{(i)}
\parallel
P_t
\right),
\end{equation}
where $P_t$ is the semantic distribution of the current validated context and $P_{t+1}^{(i)}$ is the distribution induced by the tentative update associated with candidate $y_i(t)$. This candidate-specific drift term allows the controller to penalize outputs that would move the persistent context away from the validated reasoning trajectory.
The trust score for candidate $i$ is then defined as:
\begin{equation}
T_i(t)
=
\alpha\,\mathrm{CCI}_i(t)
+
\beta\,\mathrm{AGR}_i(t)
-
\gamma\,\mathrm{CDS}_i(t),
\end{equation}
where $\alpha,\beta,\gamma>0$ control the relative importance of structural support, cross-agent agreement, and contextual drift penalty. The agreement score is computed as:
\begin{equation}
\mathrm{AGR}_i(t)
=
\frac{1}{2}
\sum_{j\neq i}
\kappa(y_i(t),y_j(t)),
\end{equation}
where $\kappa(\cdot,\cdot)$ is a semantic agreement kernel that combines entailment similarity, embedding cosine similarity, and contradiction-aware penalties. Candidates that contradict peer agents and high-confidence graph nodes receive lower agreement scores.
The controller selects the highest-trust candidate:
\begin{equation}
\hat{i}
=
\arg\max_{i\in\{1,2,3\}} T_i(t),
\qquad
\hat{y}_t
=
y_{\hat{i}}(t).
\end{equation}
This aggregation rule integrates three complementary signals. Causal consistency evaluates whether the candidate's claims are structurally supported by the validated causal graph. Cross-agent agreement reduces the influence of unreliable and compromised generations. Candidate-specific drift penalizes responses that would destabilize the persistent context if accepted. As a result, a candidate must be structurally grounded, semantically compatible with peer agents, and temporally stable to obtain high trust. This makes GT-MCP more robust than selection based on fluency, raw confidence, and majority agreement.

\subsection{Temporal Stability Objective}
\label{subsec:temporal}
GT-MCP extends per-turn trust aggregation by introducing a long-horizon temporal objective that penalizes the cumulative destabilization of the validated context. While the trust score selects the best candidate at a single turn, the temporal objective evaluates whether the sequence of selected outputs preserves structural grounding, cross-agent consistency, and contextual stability over the full interaction horizon.
Let $\hat{i}_t$ denote the selected agent index at turn $t$. The temporal stability objective is defined as:
\begin{equation}
\begin{aligned}
V_T
&=
\sum_{t=1}^{T}
\rho_T^{t-1}
\Big(
\mathrm{CCI}_{\hat{i}_t}(t)
+
\beta_T\,\mathrm{AGR}_{\hat{i}_t}(t)
-
\gamma_T\,\mathrm{CDS}_{\hat{i}_t}(t)
\\
&\qquad\qquad
-
\chi_T\,\mathbb{I}_{adv}(t)
-
\zeta_T\,\mathbb{I}_{rec}(t)
\Big).
\end{aligned}
\end{equation}
where $\rho_T\in(0,1]$ is the temporal discount factor, $\beta_T,\gamma_T,\chi_T,\zeta_T\ge0$ are weighting coefficients, $\mathbb{I}_{adv}(t)$ indicates controller-level adversarial success, and $\mathbb{I}_{rec}(t)$ indicates whether self-healing recovery is triggered at turn $t$. The term $\mathrm{CDS}_{\hat{i}_t}(t)$ denotes the candidate-specific contextual drift associated with the selected output. This objective favors trajectories in which selected outputs remain structurally supported, semantically consistent with peer agents, and close to the validated reasoning trajectory. It penalizes outputs that introduce adversarial impact, create excessive drift, and require recovery. Separating $\mathbb{I}_{adv}(t)$ from $\mathbb{I}_{rec}(t)$ is important because recovery activation does not necessarily imply a successful attack; rather, it indicates that the controller detected instability and intervened to prevent persistence. The temporal objective complements per-turn trust scoring by providing a trajectory-level criterion for evaluating controller behavior. Locally plausible responses that repeatedly introduce small deviations receive cumulative penalties through the drift and recovery terms, making gradual trajectory-steering attacks less likely to persist across turns. Thus, GT-MCP promotes stable reasoning not only by selecting high-trust outputs at each turn, but also by discouraging long-horizon accumulation of unsupported contextual changes.

\subsection{Self-Healing Recovery}
\label{subsec:heal}
When the selected output introduces excessive contextual drift, GT-MCP triggers a self-healing recovery procedure. Recovery is activated when the candidate-specific drift of the selected response exceeds the predefined threshold:
\begin{equation}
\mathrm{CDS}_{\hat{i}_t}(t) > \delta_c,
\end{equation}
where $\hat{i}_t$ denotes the selected agent at turn $t$, and $\delta_c$ is the recovery threshold. This condition ensures that recovery is triggered only when the accepted output would move the persistent context away from the validated reasoning trajectory.
Rather than resetting the entire context, GT-MCP performs targeted rollback and selective reintegration. The controller first identifies low-support claims and fragments associated with abnormal drift, then restores a prior validated checkpoint as needed. The rollback depth is chosen by minimizing a trade-off between structural risk and information loss:
\begin{equation}
d^\star
=
\arg\min_{d\in\{0,\dots,D\}}
\left(
\widehat{\mathcal{L}}(G_{t-d})
+
\lambda_R d
\right),
\end{equation}
where $D$ is the maximum rollback depth, $\widehat{\mathcal{L}}(G_{t-d})$ estimates structural vulnerability in the causal context graph at checkpoint $t-d$, and $\lambda_R>0$ penalizes excessive rollback. A larger rollback depth may remove more suspicious structure, but it may also discard validated information; the objective, therefore, selects the shallowest checkpoint that sufficiently reduces structural risk. After rollback, GT-MCP reintegrates only fragments that remain structurally grounded in the causal graph and are not linked to quarantined low-support nodes. This prevents cumulative poisoning, where an unsupported fragment accepted in one turn later becomes artificial evidence for future claims. By pruning low-support substructures while preserving validated memory, self-healing breaks this feedback loop and restores the context to a stable trajectory. Thus, self-healing functions as a trajectory-level recovery mechanism rather than a full context reset. It preserves continuity where possible, removes unsupported fragments where necessary, and maintains causal grounding, structural integrity, and trust evaluation across extended multi-turn interactions.

\subsection{Secure Multi-Agent Pipeline Algorithm}
\label{subsec:controller_alg}
The GT-MCP controller operates as a modular closed-loop pipeline that coordinates three heterogeneous LLM agents and regulates how candidate outputs update the validated context. At each turn $t$, the controller receives the validated context $c_t$ and user query $q_t$, constructs an observed context with provenance markers, obtains candidate responses from the agents, evaluates each candidate through causal consistency, cross-agent agreement, and candidate-specific contextual drift, then either accepts the selected output and triggers self-healing recovery. Algorithm~\ref{alg:gtmcp_pipeline} summarizes the complete control loop.
\begin{algorithm}[t]
\footnotesize
\caption{GT-MCP Secure Multi-Agent Context-Control Pipeline}
\label{alg:gtmcp_pipeline}
\begin{algorithmic}[1]
\Require Validated context $c_t$, query $q_t$, checkpoint store $\mathcal{H}_t$
\Require Agents $\{A_i\}_{i=1}^{3}$, trust weights $\alpha,\beta,\gamma$, recovery threshold $\delta_c$
\Ensure Selected output $\hat{y}_t$, updated context $c_{t+1}$, audit record $\mathcal{R}_t$

\State \textbf{Assemble observed context}
\State Receive untrusted inflows $\Delta c_t$
\State $\tilde{c}_t \leftarrow c_t \oplus \Delta c_t$

\State \textbf{Build structural memory}
\State $G_t \leftarrow \texttt{build\_or\_update\_graph}(c_t)$

\For{$i=1$ to $3$}
    \State \textbf{Generate candidate}
    \State $y_i(t) \leftarrow A_i(q_t,\tilde{c}_t)$

    \State \textbf{Extract and canonicalize claims}
    \State $\mathcal{S}_i \leftarrow \mathcal{E}(y_i(t))$
    \State $\mathcal{S}_i \leftarrow \texttt{canonicalize}(\mathcal{S}_i)$

    \State \textbf{Compute causal consistency}
    \State $\mathrm{CCI}_i(t) \leftarrow \texttt{causal\_consistency}(\mathcal{S}_i,G_t)$

    \State \textbf{Construct tentative update}
    \State $c_{t+1}^{(i)} \leftarrow \mathcal{U}(c_t,y_i(t))$

    \State \textbf{Compute candidate-specific drift}
    \State $\mathrm{CDS}_i(t) \leftarrow \texttt{drift}(c_{t+1}^{(i)},c_t)$
\EndFor

\For{$i=1$ to $3$}
    \State \textbf{Compute cross-agent agreement}
    \State $\mathrm{AGR}_i(t) \leftarrow \frac{1}{2}\sum_{j\neq i}\kappa(y_i(t),y_j(t))$

    \State \textbf{Aggregate trust}
    \State $T_i(t) \leftarrow \alpha\,\mathrm{CCI}_i(t)+\beta\,\mathrm{AGR}_i(t)-\gamma\,\mathrm{CDS}_i(t)$
\EndFor

\State \textbf{Select highest-trust candidate}
\State $\hat{i}_t \leftarrow \arg\max_{i\in\{1,2,3\}} T_i(t)$
\State $\hat{y}_t \leftarrow y_{\hat{i}_t}(t)$
\State $c'_{t+1} \leftarrow \mathcal{U}(c_t,\hat{y}_t)$

\State \textbf{Apply recovery gate}
\If{$\mathrm{CDS}_{\hat{i}_t}(t)>\delta_c$}
    \State $(c_{t+1},\mathcal{Q}_{t+1},d^\star) \leftarrow \texttt{self\_heal}(c'_{t+1},G_t,\mathcal{H}_t)$
    \State $\mathbb{I}_{rec}(t) \leftarrow 1$
\Else
    \State $c_{t+1} \leftarrow c'_{t+1}$
    \State $\mathcal{Q}_{t+1} \leftarrow \emptyset$
    \State $d^\star \leftarrow 0$
    \State $\mathbb{I}_{rec}(t) \leftarrow 0$
\EndIf

\State \textbf{Store audit record}
\State $\mathcal{R}_t \leftarrow \{y_i(t),\mathcal{S}_i,\mathrm{CCI}_i(t),\mathrm{AGR}_i(t),\mathrm{CDS}_i(t),T_i(t),\hat{i}_t,\hat{y}_t,\mathbb{I}_{rec}(t),\mathcal{Q}_{t+1},d^\star\}_{i=1}^{3}$

\State \Return $\hat{y}_t,c_{t+1},\mathcal{R}_t$
\end{algorithmic}
\end{algorithm}
Algorithm~\ref{alg:gtmcp_pipeline} makes the control logic explicit. The key distinction from majority voting and prompt filtering is that GT-MCP evaluates whether each candidate can safely modify persistent memory. Candidate-specific drift is computed after a tentative context update, allowing the controller to penalize responses that would destabilize future reasoning even when they appear locally fluent. If the selected candidate exceeds the recovery threshold, the self-healing mechanism restores a validated checkpoint, quarantines low-support fragments, and prevents unsupported claims from propagating across turns. The audit record preserves all candidate outputs, extracted claims, scores, selection decisions, and recovery metadata, supporting reproducibility, ablation analysis, and forensic inspection of adversarial trajectories.

\subsection{Self-Healing Algorithm}
\label{subsec:heal_alg}
The self-healing module is activated when the selected output introduces contextual drift above the recovery threshold. Its objective is to restore a stable, validated context without discarding previously grounded reasoning. Instead of resetting the entire memory, the controller identifies low-support graph regions, quarantines associated fragments, evaluates rollback candidates, and rebuilds the causal graph from the recovered context. Algorithm~\ref{alg:healing} summarizes this recovery procedure.
\begin{algorithm}[t]
\footnotesize
\caption{Context Self-Healing with Quarantine and Rollback}
\label{alg:healing}
\begin{algorithmic}[1]
\Require Tentative context $c'_{t+1}$, causal graph $G_t$, checkpoint store $\mathcal{H}_t$, maximum rollback depth $D$
\Require Structural-risk estimator $\widehat{\mathcal{L}}(\cdot)$, rollback penalty $\lambda_R$
\Ensure Healed context $c_{t+1}$, quarantine set $\mathcal{Q}_{t+1}$, rollback depth $d^\star$

\State \textbf{Identify weak graph regions}
\State $\mathcal{Z}\leftarrow \texttt{low\_support\_nodes}(G_t)$
\State $\mathcal{Z}\leftarrow \mathcal{Z}\cup \texttt{contradictory\_nodes}(G_t)$

\State \textbf{Quarantine linked fragments}
\State $\mathcal{Q}_{t+1}\leftarrow \texttt{fragments\_linked\_to}(\mathcal{Z})$
\State $\mathcal{Q}_{t+1}\leftarrow \mathcal{Q}_{t+1}\cup \texttt{unverified\_fragments}(c'_{t+1})$

\State \textbf{Prune tentative context}
\State $c^{\mathrm{prune}}\leftarrow \texttt{remove\_fragments}(c'_{t+1},\mathcal{Q}_{t+1})$

\State \textbf{Evaluate rollback candidates}
\For{$d=0$ to $D$}
    \State $G_{t-d}\leftarrow \texttt{graph\_at\_checkpoint}(\mathcal{H}_t,t-d)$
    \State $\mathrm{risk}(d)\leftarrow \widehat{\mathcal{L}}(G_{t-d})$
    \State $\mathrm{loss}(d)\leftarrow d$
    \State $\mathrm{score}(d)\leftarrow \mathrm{risk}(d)+\lambda_R\,\mathrm{loss}(d)$
\EndFor

\State \textbf{Select rollback depth}
\State $d^\star \leftarrow \arg\min_{d\in\{0,\dots,D\}}\mathrm{score}(d)$

\State \textbf{Restore checkpoint}
\State $c^{\mathrm{base}}\leftarrow \texttt{context\_at\_checkpoint}(\mathcal{H}_t,t-d^\star)$

\State \textbf{Reintegrate grounded fragments}
\State $c^{\mathrm{safe}}\leftarrow \texttt{grounded\_fragments}(c^{\mathrm{prune}},G_t,\mathcal{Q}_{t+1})$
\State $c_{t+1}\leftarrow c^{\mathrm{base}}\oplus c^{\mathrm{safe}}$

\State \textbf{Rebuild graph and update checkpoints}
\State $G_{t+1}\leftarrow \texttt{build\_causal\_graph}(c_{t+1})$
\State $\mathcal{H}_{t+1}\leftarrow \texttt{update\_checkpoints}(\mathcal{H}_t,c_{t+1},G_{t+1})$

\State \Return $c_{t+1},\mathcal{Q}_{t+1},d^\star$
\end{algorithmic}
\end{algorithm}
Algorithm~\ref{alg:healing} performs recovery at the fragment and subgraph level rather than through a full context reset. Low-support and contradictory nodes are used to identify potentially unstable fragments, while provenance links determine which parts of the tentative context should be quarantined. The rollback depth is selected by minimizing
\[
d^\star
=
\arg\min_{d\in\{0,\dots,D\}}
\left(
\widehat{\mathcal{L}}(G_{t-d})
+
\lambda_R d
\right),
\]
which balances structural risk against information loss. After restoring the selected checkpoint, GT-MCP reintegrates only fragments that remain structurally grounded and are not linked to the quarantine set. This prevents unsupported claims from becoming upstream evidence for future responses, while preserving validated memory and maintaining continuity of the reasoning trajectory.

\subsection{MCP Packet Structure and Controller Interfaces}
\label{subsec:pkt}
To operationalize context control, GT-MCP represents each interaction turn through a structured controller packet. The packet defines the information exchanged between the context manager, retrieval layer, LLM agents, trust evaluator, and recovery module. At turn $t$, the controller constructs:
\begin{equation}
\mathsf{pkt}_t
=
\left\langle
q_t,\,
c_t,\,
r_t,\,
\mathcal{P}_t,\,
\tau_t,\,
\Omega_t
\right\rangle,
\end{equation}
where $q_t$ is the user query, $c_t$ is the validated context state, $r_t$ denotes retrieved evidence and tool outputs when available, $\mathcal{P}_t$ stores provenance metadata for trusted and untrusted fragments, $\tau_t$ contains timestamps and checkpoint identifiers, and $\Omega_t$ stores controller-side configuration parameters. These parameters include trust weights $(\alpha,\beta,\gamma)$, the recovery threshold $\delta_c$, claim budget $B$, graph insertion threshold $\tau_w$, isolation flags, and recovery settings.
The packet is used to construct the observed context:
\begin{equation}
\tilde{c}_t
=
\texttt{assemble}
\left(
\mathsf{pkt}_t
\right),
\end{equation}
where \texttt{assemble} preserves the separation between validated memory, retrieved evidence, tool outputs, and untrusted inflows. This separation is essential because GT-MCP does not treat all contextual content as equally trusted. Instead, each fragment is associated with provenance metadata, confidence scores, and update permissions.
After candidate generation and trust evaluation, the controller appends decision metadata to the packet:
\begin{equation}
\mathsf{pkt}^{+}_t
=
\left\langle
\mathsf{pkt}_t,\,
\{y_i(t)\}_{i=1}^{3},\,
\{\mathcal{S}_i\}_{i=1}^{3},\,
\{T_i(t)\}_{i=1}^{3},\,
\hat{i}_t,\,
\mathcal{Q}_{t+1},\,
d^\star
\right\rangle,
\end{equation}
where $\mathcal{S}_i$ is the extracted claim set for candidate $y_i(t)$, $T_i(t)$ is its trust score, $\hat{i}_t$ is the selected agent index, $\mathcal{Q}_{t+1}$ is the quarantine set, and $d^\star$ is the rollback depth if recovery is triggered. This packetized interface makes GT-MCP auditable and reproducible. It allows the controller to trace which fragments entered the observed context, which claims were extracted, how trust scores were assigned, which output was selected, and whether recovery was activated. Therefore, the MCP packet acts as the operational bridge between multi-agent generation, causal graph validation, trust aggregation, and self-healing recovery.

\subsection{Controller Parameter Calibration}
\label{subsec:calibration}
The behavior of GT-MCP is governed by a set of controller parameters:
\begin{equation}
\theta =
\left(
\alpha,\beta,\gamma,\mu,\delta_c,\lambda_R,\tau_w,B
\right),
\end{equation}
where $\alpha,\beta,\gamma$ weight causal consistency, cross-agent agreement, and candidate-specific drift in the trust score; $\mu$ controls the stability penalty in the global objective; $\delta_c$ is the recovery threshold; $\lambda_R$ controls the rollback penalty during self-healing; $\tau_w$ is the graph insertion threshold; and $B$ is the maximum claim budget per candidate response. These parameters determine how aggressively the controller accepts candidate outputs, penalizes drift, inserts claims into the causal graph, and activates recovery. Calibration is performed over simulated adversarial interaction episodes subject to the bounded anomaly constraint:
\begin{equation}
\psi(\Delta c_t)\le \epsilon.
\end{equation}
For $N$ calibration episodes, the empirical objective is:
\begin{equation}
\begin{aligned}
\widehat{\mathcal{J}}(\theta)
&=
\frac{1}{N}
\sum_{n=1}^{N}
\sum_{t=1}^{T}
\Big[
\ell\!\left(\hat{y}_t^{(n)}\right)
+
\mu\,d\!\left(c_t^{(n)},(c_t^\star)^{(n)}\right)
\\
&\qquad\qquad
+
\gamma_C\,\mathrm{CDS}_{\hat{i}_t}^{(n)}(t)
+
\zeta_C\,\mathbb{I}_{rec}^{(n)}(t)
\Big].
\end{aligned}
\end{equation}
where $\ell(\hat{y}_t^{(n)})$ penalizes unsafe, unsupported, and adversarially influenced selected outputs; $d(c_t^{(n)},(c_t^\star)^{(n)})$ measures deviation from the ideal validated trajectory; $\mathrm{CDS}_{\hat{i}_t}^{(n)}(t)$ measures candidate-specific drift for the selected output; and $\mathbb{I}_{rec}^{(n)}(t)$ indicates whether recovery is triggered. The coefficients $\gamma_C$ and $\zeta_C$ control the calibration-level penalties for drift and recovery frequency. Because candidate selection, graph insertion, and rollback decisions are discrete and nondifferentiable, gradient-based optimization is not directly suitable. GT-MCP therefore uses gradient-free calibration, such as robust grid refinement, Bayesian optimization, and evolutionary search, depending on the available compute budget. In practice, calibration should favor parameter settings that reduce controller-level adversarial success, limit contextual drift, avoid excessive rollback, and preserve useful validated context. This calibration process balances structural grounding, agreement enforcement, drift sensitivity, and recovery behavior, producing a controller that remains stable under diverse perturbation patterns without relying solely on majority voting.

\subsection{Component-Level Evaluation and Statistical Robustness Analysis}
\label{subsec:ablation}
To assess the contribution of individual GT-MCP modules, we define ablated variants of the full controller. Let $\mathcal{M}$ denote the complete GT-MCP pipeline, and let $\mathcal{M}^{(-x)}$ denote the variant in which component $x$ is removed while the remaining controller structure is preserved. The evaluated components are causal consistency scoring (No-CCI), candidate-specific contextual drift monitoring (No-CDS), cross-agent agreement aggregation (No-AGR), and self-healing recovery (No-Heal).
For run $r$, the injection success rate is defined as:
\begin{equation}
\mathrm{ISR}^{(r)}
=
\frac{1}{T}
\sum_{t=1}^{T}
\mathbb{I}_{adv}^{(r)}(t),
\end{equation}
where $\mathbb{I}_{adv}^{(r)}(t)$ indicates controller-level adversarial success at turn $t$. To isolate the effect of component $x$, the paired difference is computed as:
\begin{equation}
\Delta_x^{(r)}
=
\mathrm{ISR}_{\mathcal{M}^{(-x)}}^{(r)}
-
\mathrm{ISR}_{\mathcal{M}}^{(r)}.
\end{equation}
The mean component effect is then:
\begin{equation}
\overline{\Delta}_x
=
\frac{1}{R}
\sum_{r=1}^{R}
\Delta_x^{(r)}.
\end{equation}

For standardized comparison, the effect size is computed as:
\begin{equation}
d_x
=
\frac{\overline{\Delta}_x}
{s_{\Delta_x}+\epsilon},
\end{equation}
where $s_{\Delta_x}$ is the sample standard deviation of the paired differences and $\epsilon$ prevents numerical instability. For non-Gaussian metrics, the same paired comparisons can be complemented with nonparametric tests, such as the Wilcoxon signed-rank test, and with rank-based effect sizes. The same evaluation logic applies to additional trajectory-level metrics, including mean causal consistency, candidate-specific contextual drift, recovery frequency, rollback depth, and utility. This component-level analysis quantifies the contributions of each GT-MCP mechanism to structural grounding, semantic stability, recovery behavior, and long-horizon robustness.

\subsection{Formal Claim Extraction and Uncertainty Modeling}
\label{subsec:claims}
GT-MCP represents each candidate output as a set of canonical atomic claims extracted from the generated text. Formally, claim extraction is defined as a controller-side mapping:
\begin{equation}
\mathcal{E}:\mathcal{Y}\rightarrow 2^{\mathcal{V}},
\end{equation}
where $\mathcal{Y}$ denotes the space of candidate outputs, $\mathcal{V}$ denotes the universe of canonical claim types, and
\begin{equation}
\mathcal{S}(y)=\mathcal{E}(y)
\end{equation}
is the extracted claim set for candidate response $y$. Because claim extraction may be affected by ambiguity, paraphrasing, incomplete grounding, and modality differences, GT-MCP models extract uncertainty probabilistically:
\begin{equation}
\mathcal{S}(y)\sim p(\mathcal{S}\mid y).
\end{equation}
Accordingly, causal consistency can be interpreted as an expectation over possible extracted claim sets:
\begin{equation}
\mathrm{CCI}_i(t)
=
\mathbb{E}_{\mathcal{S}\sim p(\mathcal{S}\mid y_i(t))}
\left[
\frac{1}{|\mathcal{S}|+\epsilon}
\sum_{v\in\mathcal{S}}
\phi(v,G_t)
\right],
\end{equation}
where $\epsilon>0$ prevents numerical instability when few and no extractable claims are available, and $\phi(v,G_t)$ denotes the bounded structural support of claim $v$ in the validated causal context graph.
To reduce manipulation through verbosity, repetition, and paraphrase inflation, GT-MCP enforces a fixed claim budget:
\begin{equation}
|\mathcal{S}(y)|\le B.
\end{equation}
Before graph insertion and trust scoring, extracted claims are canonicalized and deduplicated through entity normalization, lemmatization, modality tagging, and semantic equivalence checks. Claims that cannot be grounded, such as meta-instructions, self-referential directives, unsupported policy changes, and unverifiable tool claims, are retained for auditing but assigned a low default structural support rating. This uncertainty-aware extraction process ensures that CCI reflects genuine structural grounding rather than surface-level fluency and excessive claim volume. It also prevents adversarial candidates from artificially inflating trust scores by generating numerous redundant and semantically equivalent claims.

\section{Experimental Setup}
\label{sec:experimental_setup}
We evaluate GT-MCP in a closed-loop multi-turn adversarial setting designed to test whether malicious contextual perturbations can persist across turns, alter the validated context, and affect future reasoning. The primary evaluation horizon consists of $T=500$ interaction turns per run. Each turn contains one user query, one observed context assembly step, three candidate LLM responses, one controller selection decision, one tentative context update, and, when needed, one self-healing recovery decision. To support paired comparisons, all methods are evaluated under the same query sequence, attack schedule, and random seeds. Unless otherwise stated, each method is repeated across $R=10$ independent seeds, yielding $5{,}000$ turn-level observations per method. For paired ablation analysis, the full GT-MCP pipeline and each ablated variant are evaluated using identical seeds and identical adversarial perturbation schedules. The evaluation contains both benign and adversarial turns. Out of the $T=500$ turns in each run, $328$ turns are benign and $172$ turns contain adversarial perturbations, corresponding to $65.6\%$ benign exposure and $34.4\%$ adversarial exposure. The adversarial turns are distributed across six attack families: direct prompt injection, retrieval poisoning, tool-output injection, dormant trigger injection, gradual trajectory steering, and agreement mimicry. Table~\ref{tab:attack_schedule_setup} summarizes the turn-level evaluation schedule.
\begin{table}[t]
\centering
\caption{Turn-level evaluation schedule per run.}
\label{tab:attack_schedule_setup}
\begin{tabular}{lcc}
\toprule
Condition / Attack family & Turns & Fraction (\%) \\
\midrule
Benign interaction & 328 & 65.6 \\
Direct prompt injection & 30 & 6.0 \\
Retrieval poisoning & 40 & 8.0 \\
Tool-output injection & 30 & 6.0 \\
Dormant trigger injection & 22 & 4.4 \\
Trajectory steering & 30 & 6.0 \\
Agreement mimicry & 20 & 4.0 \\
\midrule
Total & 500 & 100.0 \\
\bottomrule
\end{tabular}
\end{table}
The GT-MCP controller uses three heterogeneous LLM agents: GPT-5.3, Llama-3.1-70B, and DeepSeek-R1. At each turn, all agents receive the same input pair $(q_t,\tilde{c}_t)$ and generate one candidate response. The controller then extracts claims, computes causal consistency, cross-agent agreement, and candidate-specific contextual drift, and selects the candidate with the highest trust score. Only the selected output is eligible to update the validated context state.
\begin{table}[t]
\centering
\caption{Controller and agent configuration.}
\label{tab:agent_controller_setup}
\begin{tabular}{p{3.1cm}p{4.7cm}}
\toprule
Component & Configuration \\
\midrule
Interaction horizon & $T=500$ turns per run \\
Repeated runs & $R=10$ seeds per method \\
Total observations & $5{,}000$ turns per method \\
LLM agents & GPT-5.3, Llama-3.1-70B, DeepSeek-R1 \\
Candidate responses & Three responses per turn \\
Selection rule & $\hat{i}_t=\arg\max_i T_i(t)$ \\
Trust signals & CCI, AGR, candidate-specific CDS \\
Recovery trigger & $\mathrm{CDS}_{\hat{i}_t}(t)>\delta_c$ \\
Context update & Only $\hat{y}_t$ updates $c_t$ through $\mathcal{U}(\cdot)$ \\
Audit logs & Candidates, claims, CCI, AGR, CDS, trust, selected agent, recovery status \\
\bottomrule
\end{tabular}
\end{table}
We compare the full GT-MCP pipeline against representative baselines and ablated variants. The baseline methods include a single-agent LLM, majority voting across agents, prompt filtering, and a retrieval-oriented defense. The ablation variants remove one GT-MCP component at a time: causal consistency scoring, cross-agent agreement, candidate-specific drift monitoring, and self-healing recovery. Table~\ref{tab:method_setup} summarizes the evaluated methods.
\begin{table}[t]
\centering
\caption{Evaluated methods and ablated variants.}
\label{tab:method_setup}
\begin{tabular}{p{3.1cm}p{4.7cm}}
\toprule
Method & Description \\
\midrule
Single-agent LLM & One LLM response is directly accepted without multi-agent control \\
Majority voting & Three agents generate responses; output is selected by semantic majority agreement \\
Prompt filtering & Candidate outputs are filtered using surface-level injection detection before update \\
RAG defense & Retrieved and contextual evidence is sanitized before candidate generation \\
No-CCI & GT-MCP without causal consistency scoring \\
No-AGR & GT-MCP without cross-agent agreement scoring \\
No-CDS & GT-MCP without candidate-specific drift monitoring \\
No-Heal & GT-MCP without rollback and quarantine recovery \\
Full GT-MCP & Complete controller with CCI, AGR, CDS, and self-healing \\
\bottomrule
\end{tabular}
\end{table}
The primary outcome is controller-level injection success rate (ISR). At turn $t$, adversarial success is recorded when the selected output belongs to $\mathcal{Y}_{adv}$, meaning that it bypasses controller constraints, inserts unsupported claims into the validated graph, manipulates tool behavior, leaks protected information, and causes persistent contextual drift without recovery. ISR is computed as:
\begin{equation}
\mathrm{ISR}
=
\frac{1}{T}
\sum_{t=1}^{T}
\mathbb{I}_{adv}(t).
\end{equation}
We also report contextual drift, causal consistency, cross-agent agreement, utility, recovery frequency, rollback depth, quarantine size, stable-turn percentage, token usage, and latency per token. Stable turns are defined as turns in which no controller-level adversarial success occurs, the selected response remains below the recovery threshold, and no persistent unsupported fragment is inserted into the validated context.
\begin{table}[t]
\centering
\caption{Evaluation metrics.}
\label{tab:metric_setup}
\begin{tabular}{p{3.0cm}p{4.8cm}}
\toprule
Metric & Definition \\
\midrule
ISR & Fraction of turns with controller-level adversarial success \\
CDS & Candidate-specific contextual drift after tentative update \\
CCI & Structural support of extracted claims in $G_t$ \\
AGR & Semantic agreement between candidate outputs \\
Utility & Per-turn controller utility combining safety and stability terms \\
Recovery frequency & Fraction of turns triggering self-healing \\
Rollback depth & Number of checkpoints reverted during recovery \\
Quarantine size & Number of fragments removed and isolated \\
Stable turns & Fraction of turns without persistent adversarial impact \\
Latency/token & Inference and controller latency normalized by token count \\
\bottomrule
\end{tabular}
\end{table}
For ablation and baseline comparisons, we compute paired differences under identical seeds. For each seed $r$, the component effect of removing module $x$ is:
\begin{equation}
\Delta_x^{(r)}
=
\mathrm{ISR}_{\mathcal{M}^{(-x)}}^{(r)}
-
\mathrm{ISR}_{\mathcal{M}}^{(r)}.
\end{equation}
We report the mean paired difference across seeds and use nonparametric paired tests when metric distributions are non-Gaussian. Turn-level metrics are summarized using the mean, median, standard deviation, percentiles, and tail-risk indicators. This setup allows us to evaluate both immediate injection failures and long-horizon trajectory instability under controlled adversarial exposure.

\section{Experimental Results}
\label{sec:results}
All evaluations are conducted under the full closed-loop GT-MCP pipeline described in Section \ref{sec:methodology}. All experiments use identical seeds across runs to ensure paired comparisons and reproducibility.

\subsection{Context Stability and Drift Behavior}
\label{subsec:drift}
\begin{figure}[t]
\centering
\includegraphics[width=\columnwidth]{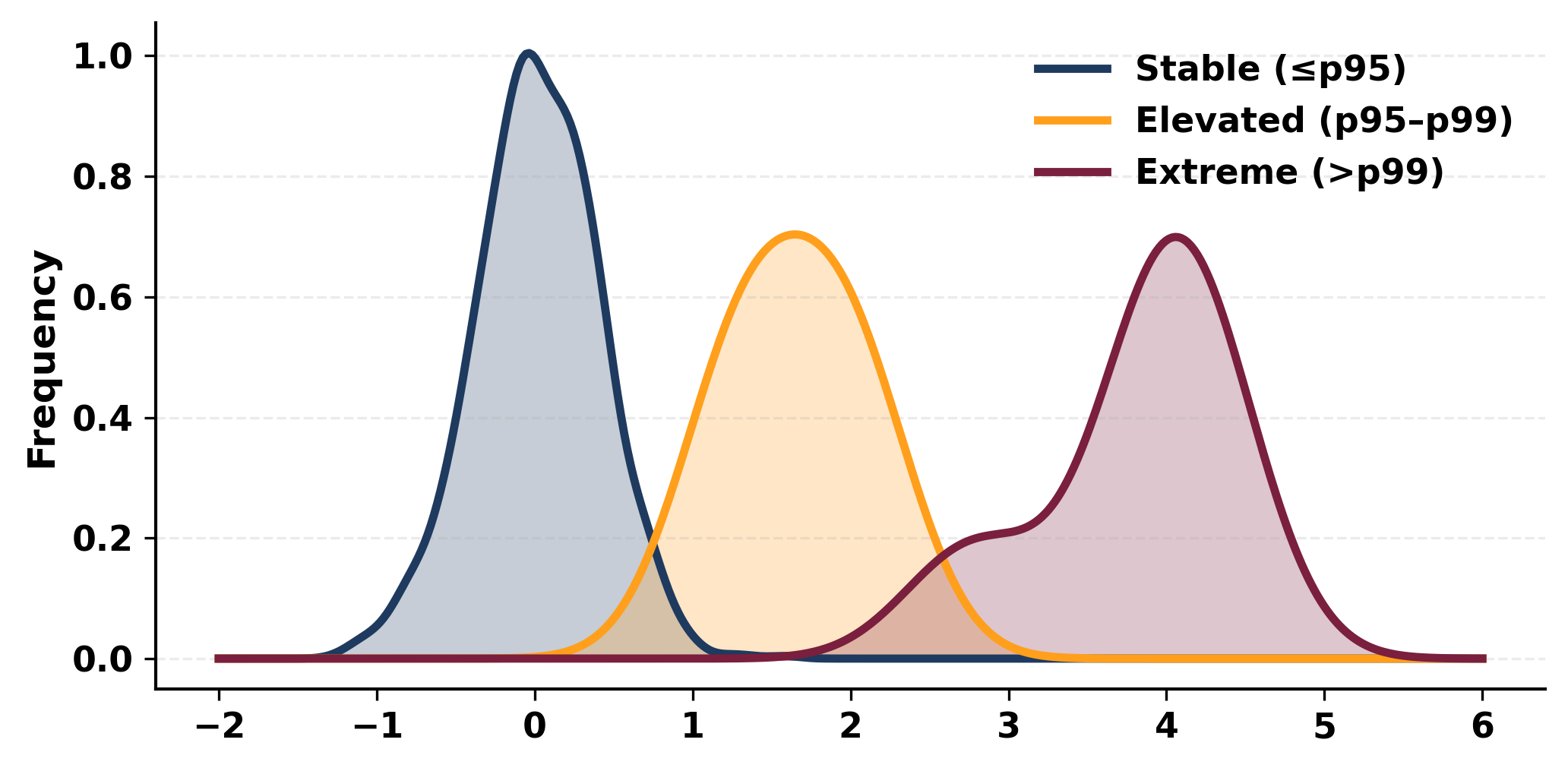}
\caption{Distribution of selected-output contextual drift across 500 interaction turns. }
\label{fig:drift_regime_dist}
\end{figure}

\begin{figure}[t]
\centering
\includegraphics[width=\columnwidth]{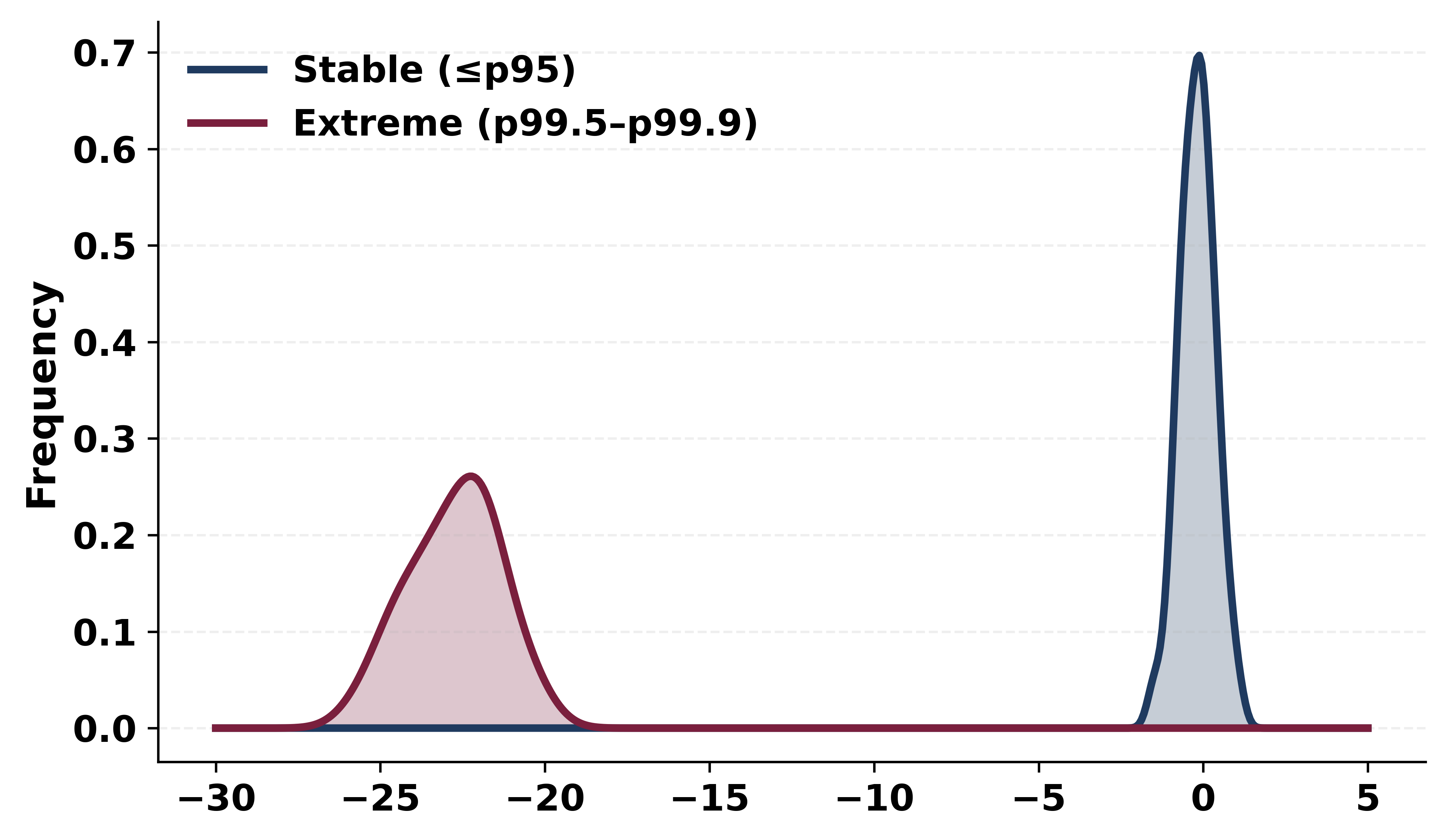}
\caption{Relationship between selected-output contextual drift and utility. }
\label{fig:drift_effect}
\end{figure}

We first examine whether GT-MCP preserves contextual stability across the $T=500$-turn closed-loop evaluation horizon defined in Section~\ref{sec:experimental_setup}. Contextual drift is measured after the controller selects a candidate response and forms the corresponding tentative context update. Therefore, the reported value corresponds to the selected-output drift $\mathrm{CDS}_{\hat{i}_t}(t)$ rather than a global input-level divergence. This metric captures whether the selected response would move the persistent context away from the validated reasoning trajectory. Near-zero drift indicates compatibility with the validated context, whereas large drift indicates a trajectory deviation that requires recovery. Figure~\ref{fig:drift_regime_dist} shows the empirical drift distribution. GT-MCP maintains near-zero drift in 498 out of 500 turns, corresponding to 99.6\% of the interaction horizon. Only two turns, representing 0.4\%, produce high-magnitude drift. This pattern demonstrates that contextual instability does not accumulate gradually over the course of the interaction. Instead, the controller keeps almost all selected outputs aligned with the validated reasoning manifold, while rare destabilizing updates are isolated as recovery-triggering events.
Table~\ref{tab:drift_compact} summarizes drift severity, utility impact, and recovery activation. The 498 near-zero-drift turns maintain a median drift of $0.000$ and a mean utility of $-0.18$, indicating stable operation around the validated trajectory. In contrast, the two high-drift turns reach a median drift of $27.63$ and a mean utility of $-23.46$, with a minimum utility of $-24.93$. Both high-drift events activate self-healing, while no recovery is triggered for near-zero-drift turns. Thus, recovery is required in only $0.4\%$ of all turns, showing that intervention is rare and reserved for severe contextual deviations.
\begin{table}[!t]
\centering
\scriptsize
\setlength{\tabcolsep}{2pt}
\caption{Drift severity, utility impact, and recovery activation.}
\label{tab:drift_compact}
\begin{tabular}{lcccccc}
\toprule
Category & Count & Fraction (\%) & Median Drift & Mean Utility & Min Utility & Heal (\%) \\
\midrule
Near-zero drift & 498 & 99.6 & 0.000 & -0.18 & -2.10 & 0.0 \\
Nonzero drift  & 2   & 0.4  & 27.63 & -23.46 & -24.93 & 100.0 \\
\bottomrule
\end{tabular}
\end{table}
Figure~\ref{fig:drift_effect} shows the relationship between drift magnitude and utility. Utility remains tightly concentrated when drift is near zero, but it collapses sharply during the two high-drift events. This confirms that utility degradation is strongly coupled to departure from the validated context trajectory. Importantly, these severe deviations remain localized because the recovery gate prevents high-drift updates from becoming persistent memory.
Table~\ref{tab:drift_regime_numeric} reports the same behavior in a compact two-block format. The absence of intermediate drift levels suggests that GT-MCP produces a sharply separated stability profile: selected outputs are either compatible with the validated context and are detected as high-risk updates requiring recovery. This separation is important for long-horizon security because gradual context poisoning typically relies on small, unsupported fragments persisting across multiple turns. In GT-MCP, such fragments fail to accumulate because unsupported updates are either assigned low trust before selection and removed through recovery after selection.
\begin{table}[!t]
\centering
\scriptsize
\caption{Drift, utility, and recovery outcomes across stable and high-drift turns.}
\label{tab:drift_regime_numeric}
\begin{tabular}{p{2.8cm} c c c c}
\toprule
Category & Count & Fraction (\%) & Median Drift & Mean Utility \\
\midrule
Near-zero drift & 498 & 99.6 & 0.00 & -0.18 \\
Nonzero drift & 2 & 0.4 & 27.63 & -23.46 \\
\bottomrule
\end{tabular}

\vspace{0.4em}

\begin{tabular}{p{2.8cm} c c}
\toprule
Category & Min Utility & Heal (\%) \\
\midrule
Near-zero drift & -2.10 & 0.0 \\
Nonzero drift & -24.93 & 100.0 \\
\bottomrule
\end{tabular}
\end{table}
Controller intervention remains minimal and localized. Table~\ref{tab:drift_extra} reports a mean drift value of $0.11$ and a median of $0.00$, confirming that high drift is exceptional rather than representative of normal operation. The mean rollback depth is $0.008$, with a median of zero, reflecting the rarity of rollback. The mean quarantine size is $1.32$, indicating that the controller isolates small sets of suspicious fragments rather than discarding large portions of context.
\begin{table}[t]
\centering
\caption{Controller intervention statistics across 500 turns.}
\label{tab:drift_extra}
\begin{tabular}{lccc}
\toprule
Metric & Mean & Median & Std \\
\midrule
Drift value & 0.11 & 0.00 & 0.90 \\
Rollback depth & 0.008 & 0.00 & 0.11 \\
Quarantine size & 1.32 & 1.00 & 0.64 \\
\bottomrule
\end{tabular}
\end{table}
Table~\ref{tab:turn_stability} summarizes percentile and worst-case behavior. The 95th percentile of drift remains zero, demonstrating that even the upper tail of ordinary operation stays aligned with the validated context. The maximum rollback depth is two checkpoints, and the maximum quarantine size is four fragments. These results show that self-healing behaves as a bounded subgraph-level correction mechanism rather than a full context reset.
\begin{table}[t]
\centering
\caption{Turn-level stability and worst-case intervention summary.}
\label{tab:turn_stability}
\begin{tabular}{lccc}
\toprule
Metric & Mean & P95 & Max \\
\midrule
Drift value & 0.11 & 0.00 & 27.63 \\
Rollback depth & 0.008 & 0.00 & 2 \\
Quarantine size & 1.32 & 2 & 4 \\
\bottomrule
\end{tabular}
\end{table}
Robustness checks confirm that this stability pattern is not an artifact of a single distance formulation. Alternative measures, including Jensen--Shannon divergence $\mathrm{CDS}_{JS}$ and 2-Wasserstein distance $\mathrm{CDS}_{W}$, reproduce the same qualitative structure: dominant near-zero drift with two isolated high-deviation events. The correlations between the operational drift score and the alternative measures remain high, exceeding $0.98$ for $\mathrm{CDS}_{JS}$ and $0.95$ for $\mathrm{CDS}_{W}$. These findings indicate that GT-MCP stabilizes context evolution by preventing unsupported fragments from becoming persistent memory updates, rather than merely filtering individual outputs.

\subsection{Multi-Agent Behavior and Selection Dynamics}
\label{subsec:agents}
\begin{figure}[t]
\centering
\includegraphics[width=\columnwidth]{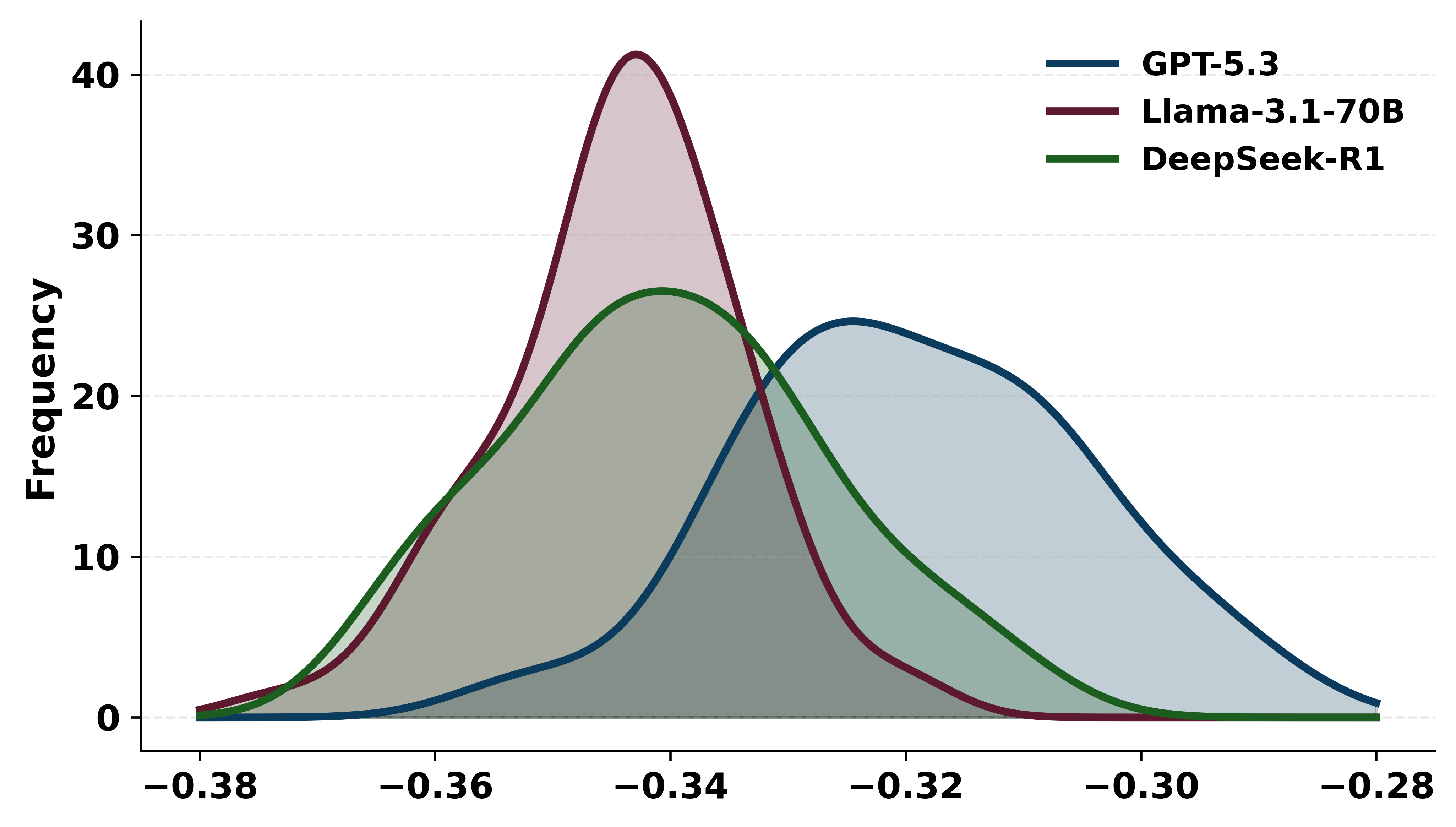}
\caption{Utility distribution of selected outputs by agent. }
\label{fig:mean_utility_agent}
\end{figure}

\begin{figure}[t]
\centering
\includegraphics[width=0.80\columnwidth]{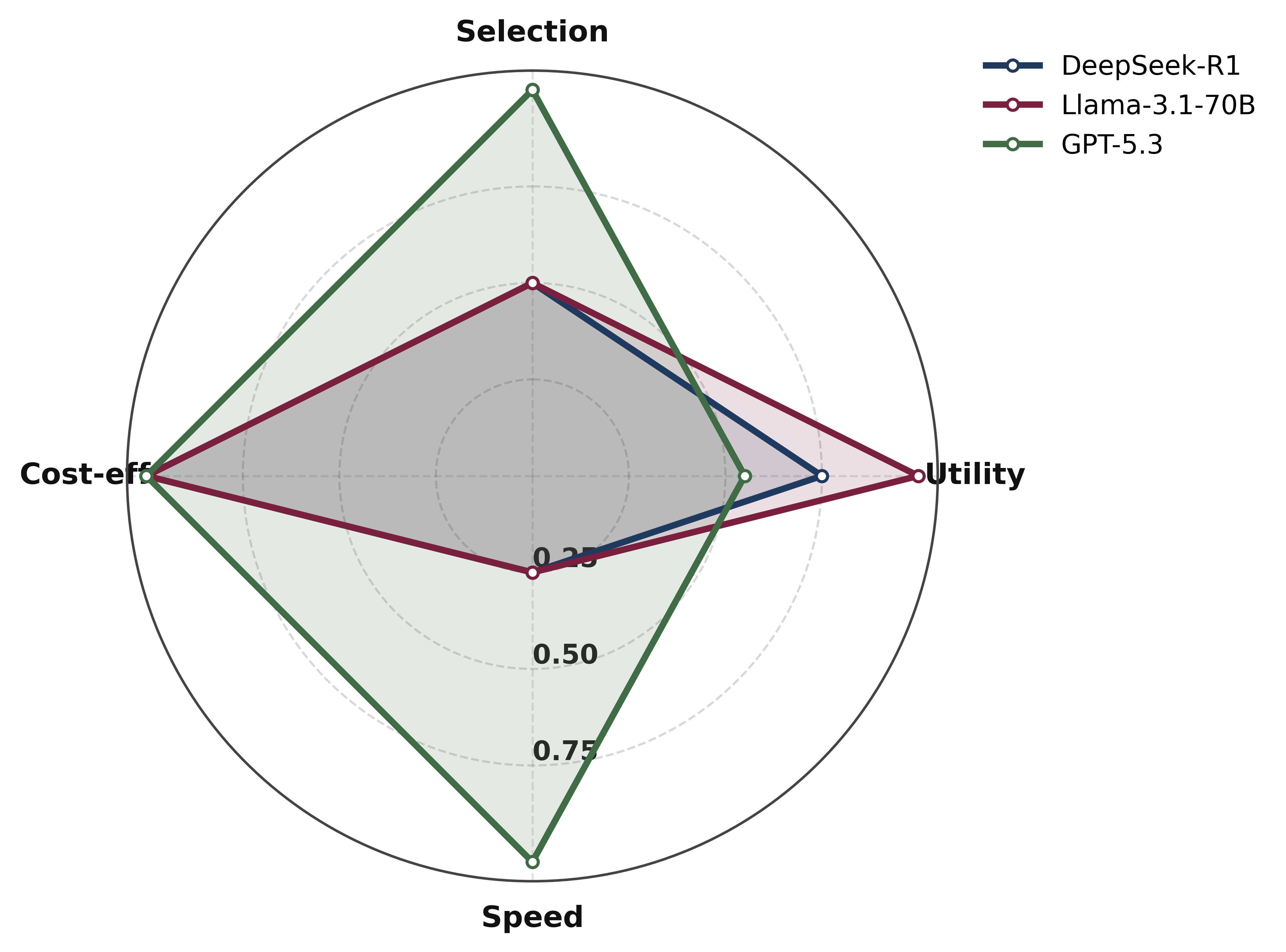}
\caption{Trust-score distributions by agent. Differences in dispersion show that the agents contribute distinct structural and semantic signals to the controller, enabling context-adaptive selection rather than fixed model preference.}
\label{fig:trust_dist}
\end{figure}

\begin{figure}[t]
\centering
\includegraphics[width=\columnwidth]{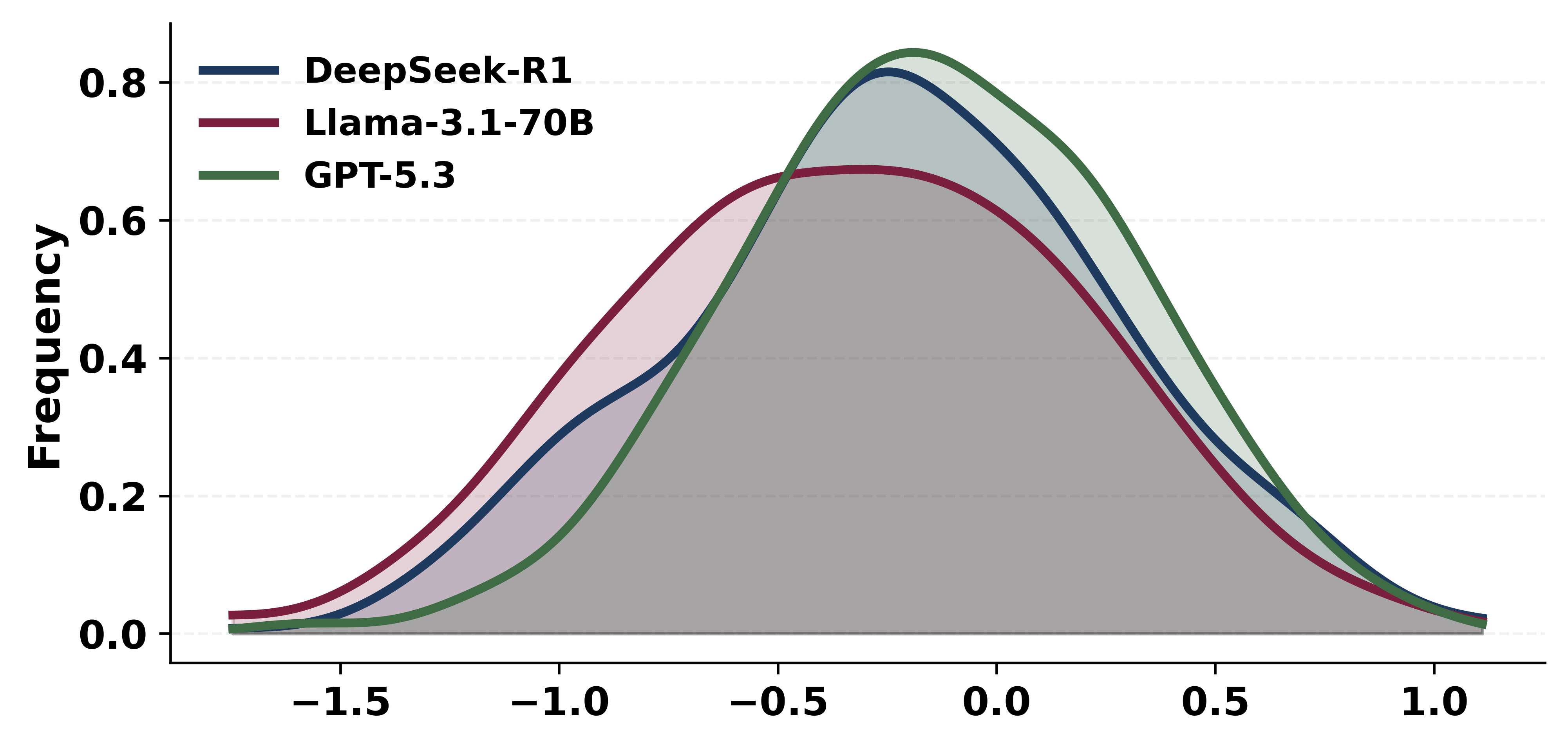}
\caption{Normalized multi-objective profile of the three LLM agents. }
\label{fig:radar_agents}
\end{figure}

We next analyze how the three heterogeneous LLM agents contribute to GT-MCP selection behavior across the $T=500$-turn evaluation horizon. The evaluated agents are GPT-5.3, Llama-3.1-70B, and DeepSeek-R1. All agents receive the same observed input pair $(q_t,\tilde{c}_t)$ at each turn, but their candidate responses are selected only after controller-side evaluation through causal consistency, cross-agent agreement, and candidate-specific contextual drift. This design allows GT-MCP to exploit model diversity without assuming that any single agent is consistently superior.
Figure~\ref{fig:mean_utility_agent} shows the utility distribution associated with selected outputs from each agent. The distributions overlap substantially, indicating that the controller does not operate as a static model-ranking mechanism. Instead, selection depends on context-specific alignment with the validated causal graph and the current reasoning trajectory. GPT-5.3 and Llama-3.1-70B show wider utility dispersion because they are selected more frequently and therefore face broader contextual exposure. DeepSeek-R1 has lower variance, suggesting a stabilizing role in cases where structurally conservative responses are preferred.
The controller assigns trust according to:
\[
T_i(t)
=
\alpha\,\mathrm{CCI}_i(t)
+
\beta\,\mathrm{AGR}_i(t)
-
\gamma\,\mathrm{CDS}_i(t),
\]
where $\mathrm{CDS}_i(t)$ is the candidate-specific drift induced by the tentative update associated with agent $i$. This formulation is important because drift must affect the ranking of candidate outputs; a global drift term shared by all agents would not influence the selection decision. Trust, therefore, reflects whether a candidate is structurally grounded, semantically compatible with peer agents, and safe to integrate into a persistent context.
Figure~\ref{fig:trust_dist} reports the trust-score distributions. Llama-3.1-70B exhibits the widest trust dispersion, indicating sensitivity to graph structure and contextual variation. GPT-5.3 shows moderate trust variance with the highest selection frequency, suggesting frequent alignment with the controller's combined structural and semantic criteria. DeepSeek-R1 shows the narrowest distribution, consistent with stable but less frequently selected behavior. This heterogeneity is beneficial: it creates separable trust profiles that allow the controller to avoid reducing multi-agent selection to simple voting.
Figure~\ref{fig:radar_agents} summarizes the multi-objective agent profiles, including selection frequency, latency, utility, structural stability, and cost efficiency. GPT-5.3 contributes the highest selection frequency and strong utility alignment, Llama-3.1-70B provides broader variability and fast responses in favorable contexts, and DeepSeek-R1 contributes the most stable low-variance profile. The key observation is that GT-MCP benefits from controlled heterogeneity: agents differ enough to provide useful alternatives, but selection remains constrained by causal grounding and drift control.
Table~\ref{tab:agent_numeric_small} provides quantitative support for these observations. GPT-5.3 is selected in 50.0\% of turns, followed by Llama-3.1-70B in 33.8\% and DeepSeek-R1 in 16.2\%. Despite this imbalance in selection frequency, the mean CCI and AGR values remain close across agents. This indicates that selection is not driven by raw agent identity, but by local compatibility with the validated context. DeepSeek-R1 has the least negative mean utility and the lowest utility variance, supporting its role as a conservative stabilizing contributor.
\begin{table}[!t]
\centering
\caption{Per-agent utility, selection frequency, and structural scores.}
\label{tab:agent_numeric_small}

\scriptsize
\begin{tabular}{l c c c c}
\toprule
Agent & Mean Utility & Median Utility & Std & Sel(\%) \\
\midrule
GPT-5.3        & -0.331 & -0.201 & 1.60 & 50.0 \\
Llama-3.1-70B  & -0.318 & -0.178 & 1.71 & 33.8 \\
DeepSeek-R1    & -0.131 & -0.171 & 0.37 & 16.2 \\
\bottomrule
\end{tabular}

\vspace{0.5em}

\begin{tabular}{l c c}
\toprule
Agent & Mean CCI & Mean AGR \\
\midrule
GPT-5.3        & 0.103 & 0.428 \\
Llama-3.1-70B  & 0.097 & 0.432 \\
DeepSeek-R1    & 0.092 & 0.424 \\
\bottomrule
\end{tabular}

\end{table}
Table~\ref{tab:agent_risk_small} reports instability indicators. Severe utility degradation and drift-triggered recovery occur only for GPT-5.3 and Llama-3.1-70B, with one event each across the 500-turn horizon. This corresponds to 0.2\% per affected agent and 0.4\% in total, matching the overall recovery frequency reported in Section~\ref{subsec:drift}. DeepSeek-R1 shows no severe degradation, no drift-trigger event, and no post-selection recovery. These results suggest that risk exposure is associated with selection frequency and contextual coverage, rather than with inherent model unreliability.
\begin{table}[!t]
\centering
\scriptsize
\caption{Agent-level instability indicators across the 500-turn horizon.}
\label{tab:agent_risk_small}
\begin{tabular}{l c c c}
\toprule
Agent & Utility $<-1$ (\%) & Drift-trigger (\%) & Heal-after-selection (\%) \\
\midrule
GPT-5.3        & 0.2 & 0.2 & 0.2 \\
Llama-3.1-70B  & 0.2 & 0.2 & 0.2 \\
DeepSeek-R1    & 0.0 & 0.0 & 0.0 \\
\bottomrule
\end{tabular}
\end{table}
Selection confidence is summarized in Table~\ref{tab:agent_confidence}. Mean trust gaps remain below 0.05 for all agents, showing that GT-MCP often operates in a narrow decision band rather than selecting trivially dominant candidates. Stable win rates exceed 98\% for all agents, confirming that selected outputs usually remain aligned with the validated context and do not require recovery. DeepSeek-R1 has the smallest mean trust gap but the highest stable win rate, reflecting conservative selection under highly stable conditions.
\begin{table}[!t]
\centering
\caption{Selection confidence and stable-win behavior.}
\label{tab:agent_confidence}
\begin{tabular}{lccc}
\toprule
Agent & Mean Trust Gap & Win Margin & Stable Win (\%) \\
\midrule
GPT-5.3        & 0.042 & 0.018 & 98.8 \\
Llama-3.1-70B  & 0.039 & 0.017 & 98.6 \\
DeepSeek-R1    & 0.028 & 0.011 & 99.4 \\
\bottomrule
\end{tabular}
\end{table}
Additionally, these results show that GT-MCP converts heterogeneous model behavior into coordinated context control. The controller does not merely choose the most fluent and most frequently agreeing model; it selects the response that best satisfies structural grounding, peer consistency, and candidate-specific drift constraints. Stability, therefore, emerges from controlled model diversity and trust-based regulation of context updates, rather than from reliance on a single dominant LLM.

\subsection{Global Performance Distribution}
\label{subsec:global_utility}
\begin{figure}[t]
\centering
\includegraphics[width=\columnwidth]{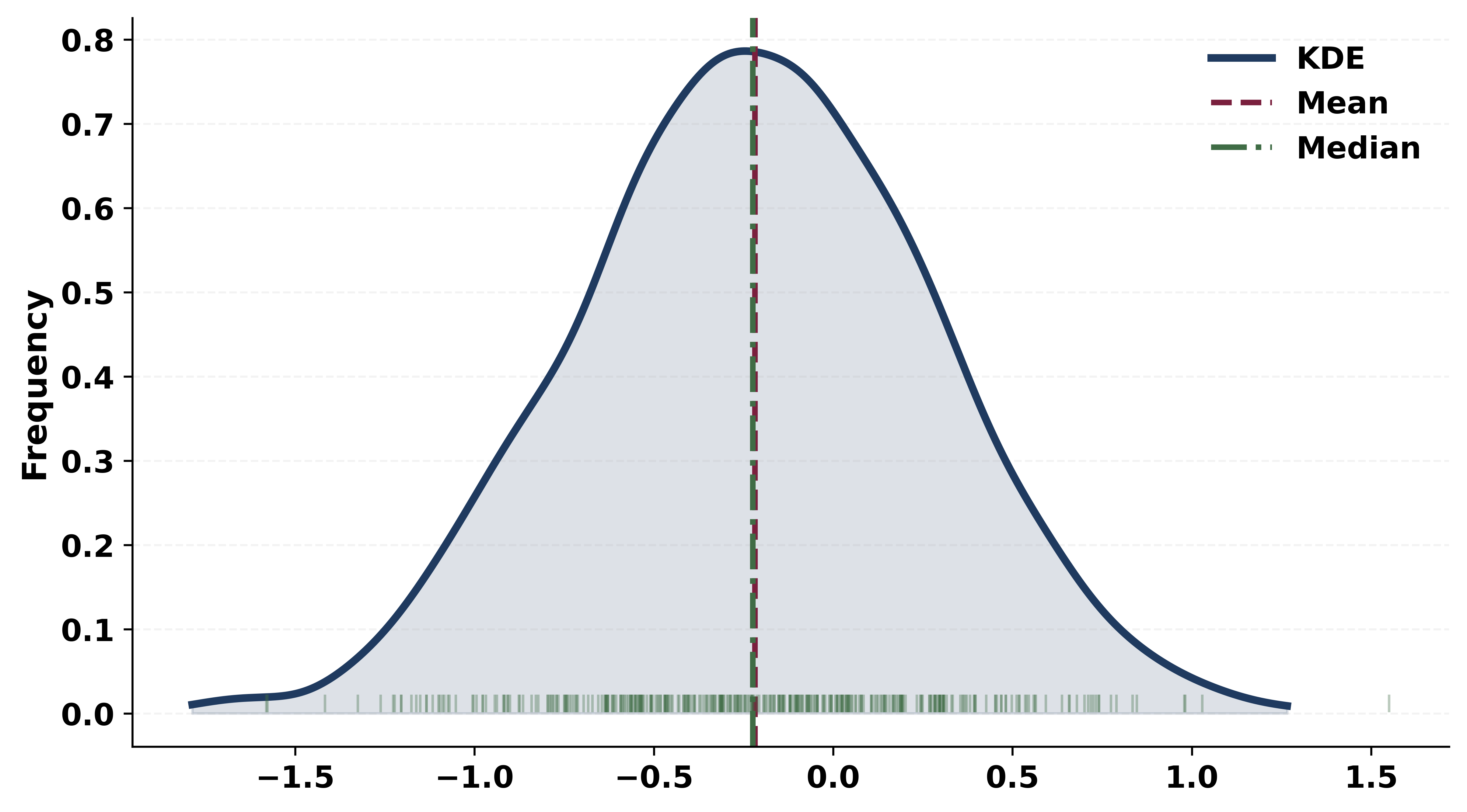}
\caption{Distribution of per-turn utility across the 500-turn GT-MCP evaluation. }
\label{fig:global_utility_dist}
\end{figure}

We characterize long-horizon performance by analyzing the marginal distribution of per-turn utility across the full GT-MCP pipeline. In contrast to single-output evaluation, this analysis captures the cumulative effect of repeated trust-based selection, causal validation, candidate-specific drift control, and self-healing recovery over the $T=500$-turn evaluation horizon. Figure~\ref{fig:global_utility_dist} shows that utility is strongly concentrated around a compact central range, with a median of $-0.19$ and a 90\% interval from $-0.72$ to $0.30$. This concentration indicates that typical GT-MCP operation remains stable and predictable, even under adversarial exposure.
The distribution is left-skewed because the only severe deviations occur in the harmful tail. As shown in Table~\ref{tab:global_utility_numeric}, the minimum utility reaches $-24.93$, whereas the upper tail remains bounded with a maximum of $0.78$. This asymmetry is consistent with the drift analysis in Section~\ref{subsec:drift}: severe utility degradation is not a persistent operating state, but a rare event associated with high contextual drift. The central utility mass remains compact, indicating that recovery events do not distort the controller's dominant long-horizon behavior.
\begin{table}[!t]
\centering
\caption{Global utility distribution statistics.}
\label{tab:global_utility_numeric}
\begin{tabular}{lc}
\toprule
Statistic & Value \\
\midrule
Min & -24.93 \\
P05 & -0.72 \\
Median & -0.19 \\
P95 & 0.30 \\
Max & 0.78 \\
\bottomrule
\end{tabular}
\end{table}
Table~\ref{tab:utility_tail} confirms that tail risk is highly localized. Only two turns fall below each severe utility threshold, corresponding to $0.4\%$ of the evaluation horizon. Because the same two events account for all thresholds below $-1$, $-5$, and $-10$, the results indicate isolated severe shocks rather than repeated moderate degradation. This is important for trajectory-level security: adversarial perturbations may create rare local failures, but they do not produce sustained deterioration of the validated context trajectory.
\begin{table}[t]
\centering
\caption{Tail risk indicators for severe utility degradation.}
\label{tab:utility_tail}
\begin{tabular}{lcc}
\toprule
Threshold & Count & Fraction (\%) \\
\midrule
Utility $<-1$ & 2 & 0.4 \\
Utility $<-5$ & 2 & 0.4 \\
Utility $<-10$ & 2 & 0.4 \\
\bottomrule
\end{tabular}
\end{table}
The dispersion indicators in Table~\ref{tab:utility_dispersion} further support this interpretation. The interquartile range is only $0.41$, showing that normal operation is tightly bound. The negative skewness reflects the rare harmful tail, while the elevated kurtosis indicates that extreme deviations are concentrated in a small number of observations rather than spread across the full interaction horizon. Thus, GT-MCP exhibits a stable central performance profile with sparse, recoverable tail events.
\begin{table}[!t]
\centering
\caption{Utility dispersion indicators across the full evaluation horizon.}
\label{tab:utility_dispersion}
\begin{tabular}{lc}
\toprule
Metric & Value \\
\midrule
Variance & 0.169 \\
IQR & 0.41 \\
Skewness & -3.8 \\
Kurtosis & 21.4 \\
\bottomrule
\end{tabular}
\end{table}
Table~\ref{tab:global_effectsize_small} summarizes the distributional robustness interpretation without introducing unsupported placeholder statistics. The utility profile is dominated by a single compact central mass, central dispersion remains bounded, and severe tail events are limited to $0.4\%$ of turns. Together, these diagnostics show that GT-MCP maintains stable utility under normal operation while containing adversarially induced degradation to isolated events that do not reshape the long-horizon performance distribution.
\begin{table}[t]
\centering
\scriptsize
\setlength{\tabcolsep}{2pt}
\renewcommand{\arraystretch}{1.15}
\caption{Global distribution robustness diagnostics.}
\label{tab:global_effectsize_small}
\begin{tabularx}{\columnwidth}{p{2.1cm} p{2.5cm} X}
\toprule
Diagnostic & Observation & Interpretation \\
\midrule
Central concentration & Median $=-0.19$, IQR $=0.41$ & Stable typical behavior \\
Tail localization & Severe events $=0.4\%$ & Rare isolated degradation \\
Left-tail asymmetry & Skewness $=-3.8$ & Harmful shocks are sparse \\
Tail sharpness & Kurtosis $=21.4$ & Extremes are concentrated, not persistent \\
Upper-tail bound & P95 $=0.30$, Max $=0.78$ & Positive deviations remain limited \\
\bottomrule
\end{tabularx}
\end{table}
In addition, the global utility distribution demonstrates that GT-MCP produces a stable long-horizon performance profile. The controller keeps most turns within a compact utility band, while severe degradation is confined to two high-drift events that are handled by the recovery mechanism. This supports the central claim that GT-MCP stabilizes contextual reasoning not by eliminating every perturbation, but by preventing rare destabilizing events from becoming persistent context corruption.

\subsection{Latency and Compute Efficiency}
\label{subsec:efficiency}
\begin{figure}[t]
\centering
\includegraphics[width=\columnwidth]{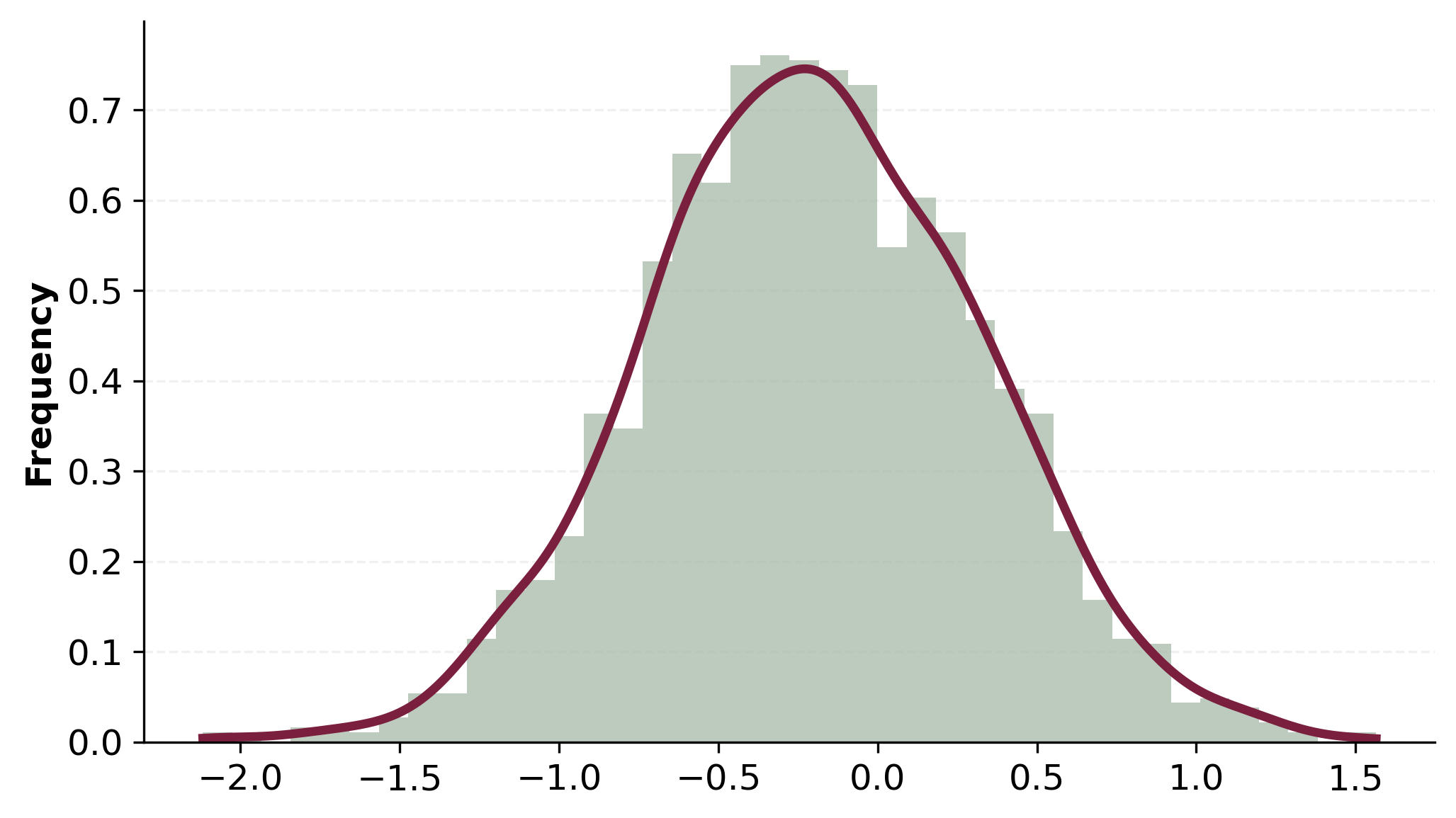}
\caption{Latency--utility relationship across 500 interaction turns. Utility remains concentrated across latency levels, indicating that increased processing time does not substantially degrade the quality of selected output.}
\label{fig:latency_utility}
\end{figure}

\begin{figure}[!t]
\centering
\includegraphics[width=\columnwidth]{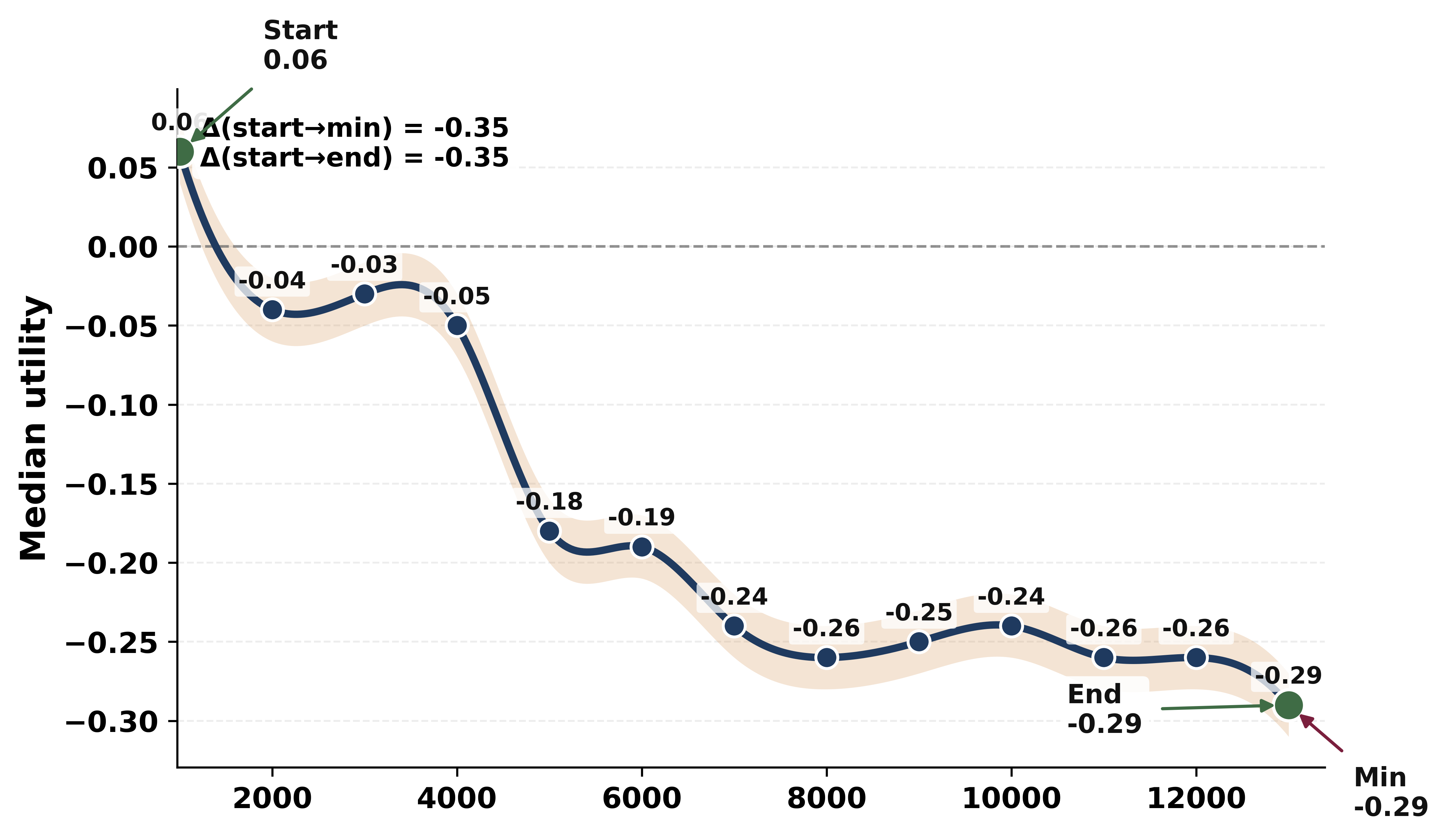}
\caption{Token usage versus median utility. }
\label{fig:compute_utility}
\end{figure}

We evaluate whether the stability gains of GT-MCP are achieved with a predictable computational cost. Figure~\ref{fig:latency_utility} shows the relationship between latency and per-turn utility across the $T=500$-turn evaluation horizon. Utility remains concentrated within the same compact band observed in Section~\ref{subsec:global_utility}, while the latency-utility trend shows only a weak negative slope. This indicates that additional inference and controller processing do not amplify instability; instead, the closed-loop pipeline preserves validated context alignment across a wide range of latency values.
Table~\ref{tab:latency_eff} summarizes latency-normalized indicators. Latency per token remains tightly concentrated, with a mean of $0.00163$\,s and a standard deviation of $0.00021$\,s. Utility per second remains close to zero, consistent with the earlier bounded utility distribution. Stability per second remains positive, suggesting that additional computation is primarily directed toward structural validation and agreement checking rather than unstable output expansion.
\begin{table}[t]
\centering
\caption{Latency-normalized efficiency indicators.}
\label{tab:latency_eff}
\begin{tabular}{lccc}
\toprule
Metric & Mean & Median & Std \\
\midrule
Latency per token & 0.00163 & 0.00159 & 0.00021 \\
Utility per second & -0.0092 & -0.0071 & 0.012 \\
Stability per second & 0.0041 & 0.0038 & 0.002 \\
\bottomrule
\end{tabular}
\end{table}
Token-level scaling is shown in Figure~\ref{fig:compute_utility}. Median utility decreases mildly as token usage increases, but the curve flattens at higher token counts. This behavior is consistent with diminishing marginal utility rather than instability amplification. In other words, longer generations do not lead to a proportional increase in contextual risk because the controller filters out unsupported claims before they can update persistent memory.
Table~\ref{tab:token_regimes_compact} partitions token usage into three approximately equal groups. Median utility shifts from $-0.01$ in the lower-token group to $-0.22$ in the higher-token group, while latency increases only from $31.4$\,s to $32.6$\,s. The small utility shift and narrow latency range indicate predictable scaling: higher token usage slightly increases processing cost but does not introduce threshold-driven degradation.
\begin{table}[t]
\centering
\scriptsize
\caption{Token usage groups and associated performance indicators.}
\label{tab:token_regimes_compact}
\begin{tabular}{lcc}
\toprule
Category & Fraction (\%) & Median Tokens \\
\midrule
Lower token usage & 33.4 & 7,814 \\
Moderate token usage & 33.2 & 20,008 \\
Higher token usage & 33.4 & 32,019 \\
\bottomrule
\end{tabular}

\vspace{0.5em}

\begin{tabular}{lcc}
\toprule
Category & Median Utility & Latency (s) \\
\midrule
Lower token usage & -0.01 & 31.4 \\
Moderate token usage & -0.19 & 32.2 \\
Higher token usage & -0.22 & 32.6 \\
\bottomrule
\end{tabular}
\end{table}
Normalized compute indicators in Table~\ref{tab:compute_eff} show that utility per token remains close to zero, whereas CCI and AGR per token remain positive and stable. This indicates that additional tokens contribute more to structural grounding and inter-agent agreement signals than to marginal utility gain. Moreover, this is desirable: increased reasoning depth strengthens validation evidence without materially widening the utility distribution.
\begin{table}[t]
\centering
\caption{Normalized compute efficiency indicators.}
\label{tab:compute_eff}
\begin{tabular}{lccc}
\toprule
Metric & Mean & Median & Std \\
\midrule
Utility per token & $-1.4{\times}10^{-5}$ & $-9.1{\times}10^{-6}$ & $2.3{\times}10^{-5}$ \\
CCI per token & $5.1{\times}10^{-6}$ & $4.7{\times}10^{-6}$ & $1.2{\times}10^{-6}$ \\
AGR per token & $2.1{\times}10^{-5}$ & $2.0{\times}10^{-5}$ & $3.2{\times}10^{-6}$ \\
\bottomrule
\end{tabular}
\end{table}
Table~\ref{tab:efficiency_effectsize_compact} summarizes the scaling interpretation without relying on placeholder statistical estimates. The observed relationship between token count and utility is weak, the marginal utility cost remains small, and latency does not independently drive degradation after accounting for token usage. These results indicate that GT-MCP scales predictably: additional computation increases validation depth while preserving structural consistency and bounded utility dispersion.
\begin{table}[!t]
\centering
\tiny
\setlength{\tabcolsep}{1.2pt}
\renewcommand{\arraystretch}{1.12}
\caption{Efficiency scaling diagnostics.}
\label{tab:efficiency_effectsize_compact}
\begin{tabularx}{\columnwidth}{
>{\raggedright\arraybackslash}p{1.75cm}
>{\raggedright\arraybackslash}p{2.25cm}
>{\raggedright\arraybackslash}X
}
\toprule
Diagnostic & Observation & Interpretation \\
\midrule
Token--utility association & Weak negative trend & No strong monotonic degradation \\
Marginal utility cost & Small across token groups & Bounded cost of longer outputs \\
Latency effect & Near-flat latency--utility curve & Runtime alone does not drive failure \\
Validation density & Positive CCI/token and AGR/token & Extra tokens support verification signals \\
Scaling behavior & Smooth across token groups & No threshold-driven collapse \\
\bottomrule
\end{tabularx}
\end{table}
Furthermore, GT-MCP maintains predictable compute--performance behavior. Longer outputs and higher latency introduce mild efficiency costs, but they do not destabilize the reasoning trajectory. The controller converts additional computation into causal validation and inter-agent consistency checks, preventing token growth from becoming a channel for cumulative contextual drift.

\subsection{Reasoning Structure and Manifold Properties}
\label{subsec:manifold}
\begin{figure}[!t]
\centering
\includegraphics[width=\columnwidth]{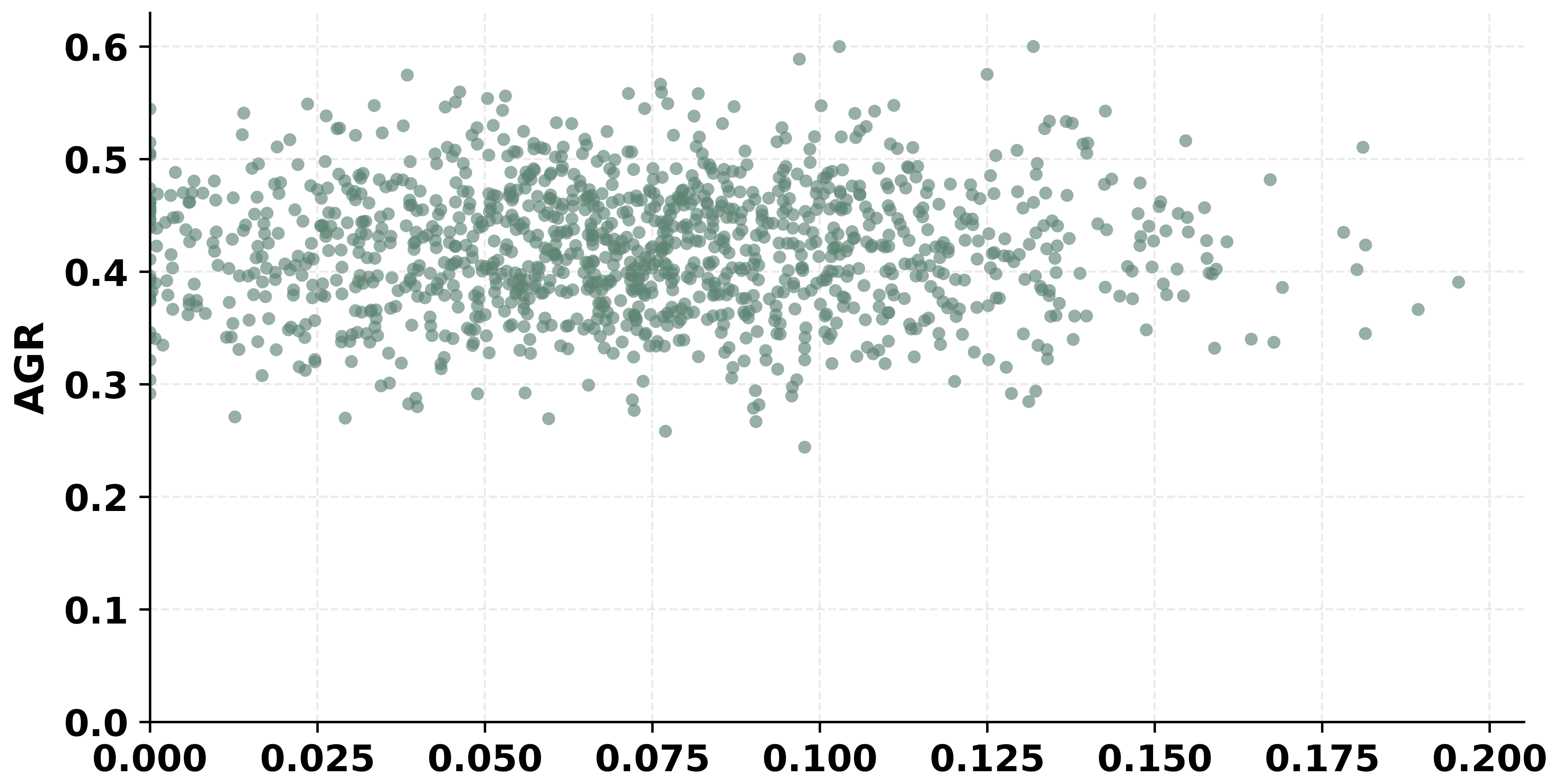}
\caption{Empirical reasoning space defined by causal consistency and cross-agent agreement. }
\label{fig:reasoning_space}
\end{figure}

\begin{figure}[t]
\centering
\includegraphics[width=\columnwidth]{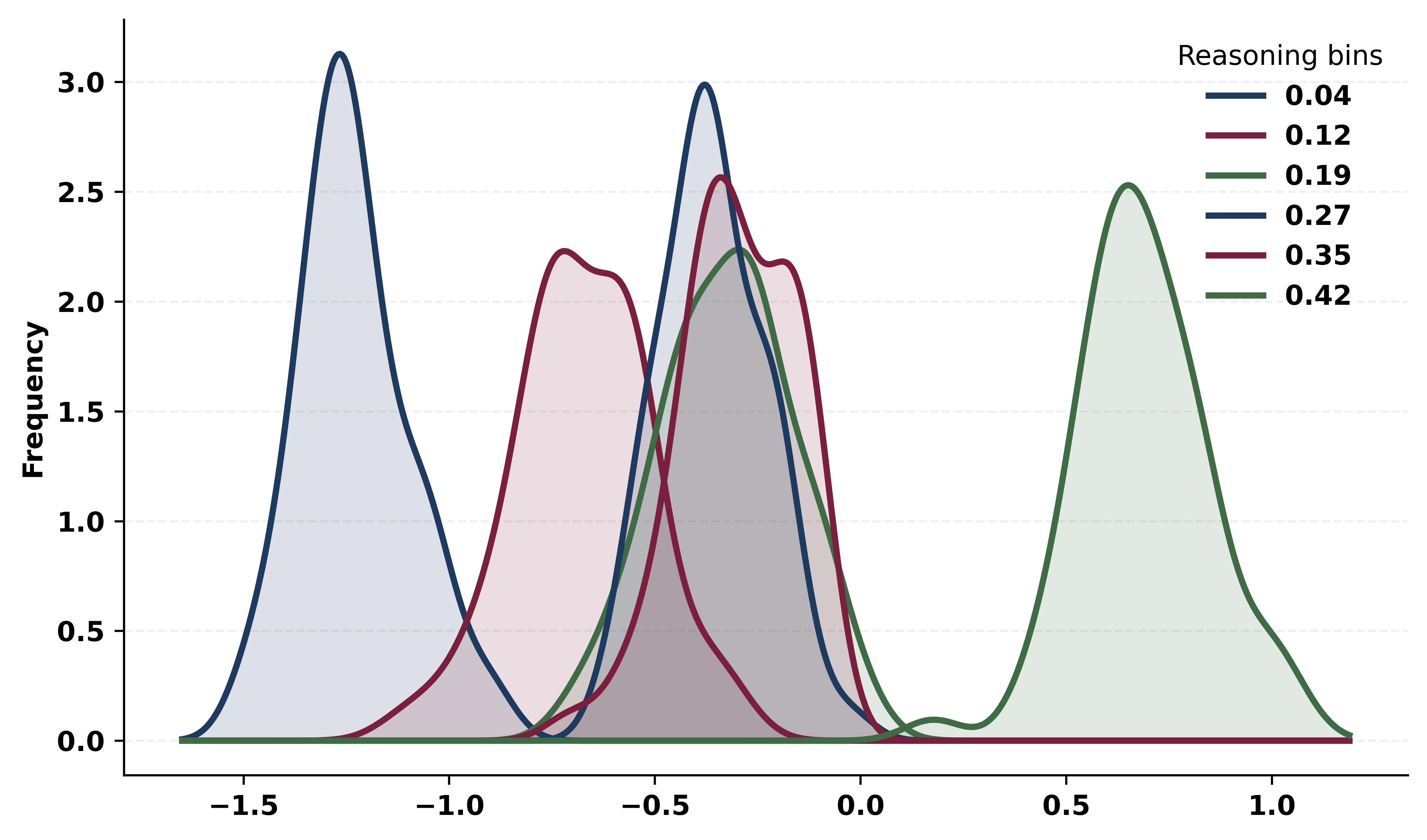}
\caption{Manifold alignment score versus utility. }
\label{fig:manifold_utility}
\end{figure}

We next examine whether GT-MCP merely filters unsafe outputs and whether it shapes the structure of the reasoning space itself. The controller evaluates each selected response through causal consistency, cross-agent agreement, and candidate-specific drift. As a result, accepted outputs are expected to occupy a constrained region of the semantic space where claims are structurally grounded and mutually coherent.
Figure~\ref{fig:reasoning_space} shows the joint distribution of CCI and AGR. The points concentrate within a narrow band rather than spreading across the full CCI-AGR plane. This indicates that GT-MCP does not accept arbitrary fluent outputs; instead, it favors responses that satisfy both graph-based support and inter-agent semantic compatibility. In this sense, the controller induces a structured reasoning manifold: a restricted subset of candidate responses that remain compatible with the validated context trajectory.
Table~\ref{tab:structural_metrics} quantifies this structure. CCI and AGR remain bound with moderate dispersion across the full evaluation horizon. The trust score has a negative mean because it incorporates drift penalties and serves as a relative selection potential rather than an absolute measure of correctness. Therefore, its value should be interpreted comparatively across candidates at each turn, not as a standalone quality measure. The bounded dispersion of CCI and AGR supports the observation that GT-MCP repeatedly selects outputs from a stable structural region of the reasoning space.
\begin{table}[t]
\centering
\caption{Structural reasoning metrics across all interaction turns.}
\label{tab:structural_metrics}
\begin{tabular}{lccc}
\toprule
Metric & Mean & Median & Std \\
\midrule
CCI   & 0.098 & 0.091 & 0.052 \\
AGR   & 0.419 & 0.420 & 0.062 \\
Trust & -0.294 & -0.187 & 0.411 \\
\bottomrule
\end{tabular}
\end{table}
Figure~\ref{fig:manifold_utility} links this structural organization to performance. Utility increases as manifold alignment improves, showing that responses closer to the structurally grounded region tend to produce better outcomes. Low-alignment responses are associated with negative utility, whereas higher-alignment responses compress outcomes toward neutral and positive values. This pattern indicates that the manifold is not only a geometric artifact of the scoring function; it is predictive of downstream utility.
Table~\ref{tab:manifold_stability} reports percentile-level structural stability. The 95th percentile values of CCI and AGR remain bounded, and the trust score increases smoothly across percentiles. This suggests that GT-MCP performs continuous trust-weighted selection rather than abrupt threshold-based switching. The controller, therefore, stabilizes reasoning by gradually favoring candidates that are more structurally supported, more semantically consistent, and less likely to destabilize the validated context.
\begin{table}[!t]
\centering
\scriptsize
\caption{Percentile-level stability of structural reasoning indicators.}
\label{tab:manifold_stability}
\begin{tabular}{lccc}
\toprule
Metric & 50th percentile & 75th percentile & 95th percentile \\
\midrule
CCI   & 0.091 & 0.135 & 0.198 \\
AGR   & 0.420 & 0.470 & 0.502 \\
Trust & -0.187 & 0.050 & 0.320 \\
\bottomrule
\end{tabular}
\end{table}
Table~\ref{tab:manifold_effectsize} further quantifies the relationship between manifold alignment and utility. The Spearman correlation of $0.82$ indicates monotonic coupling, while the Theil-Sen slope of $0.47$ shows a positive robust trend even under heavy-tailed utility behavior. Cliff's $\delta=0.68$ indicates a substantial separation between low- and high-alignment outputs. Together, these diagnostics show that alignment with the structural reasoning manifold is strongly associated with improved controller outcomes.
\begin{table}[!t]
\centering
\scriptsize
\caption{Manifold-utility coupling diagnostics.}
\label{tab:manifold_effectsize}
\begin{tabular}{lc}
\toprule
Metric & Estimate \\
\midrule
Spearman $\rho$(manifold, utility) & 0.82 \\
Theil--Sen slope & 0.47 \\
Cliff's $\delta$ (low vs high bins) & 0.68 \\
\bottomrule
\end{tabular}

\vspace{0.5em}

\begin{tabular}{lc}
\toprule
Interpretation & Purpose \\
\midrule
Strong monotonic association & Alignment--utility coupling \\
Positive robust slope & Marginal utility gain \\
Large practical separation & Low vs. high alignment effect \\
\bottomrule
\end{tabular}
\end{table}
Moreover, these results show that GT-MCP shapes the reasoning process into a constrained structural subspace rather than simply filtering individual outputs. Responses aligned with this subspace achieve higher utility, while responses farther from it are penalized through lower trust and increased drift risk. This supports the trajectory-level interpretation of GT-MCP: stability arises from maintaining reasoning within validated structural bounds, not merely from rejecting isolated unsafe generations.

\subsection{Game-Theoretic Outcomes}
\label{subsec:game_outcomes}
We evaluate whether the trust-based controller changes the strategic payoff of adversarial perturbation. In GT-MCP, the controller commits to a selection policy based on causal consistency, cross-agent agreement, and candidate-specific contextual drift, while the attacker attempts to induce selected outputs that corrupt the validated context. The key question is therefore not only whether attacks reduce utility but also whether they create a persistent advantage for the attacker over the interaction horizon.
Table~\ref{tab:game_numeric_small} summarizes outcomes under benign and adversarial exposure. The evaluation contains 328 benign turns and 172 adversarial turns, matching the exposure schedule defined in Section~\ref{sec:experimental_setup}. Adversarial turns show lower mean utility than benign turns ($-0.56$ versus $-0.16$), indicating that attacks increase local difficulty. However, the attacker's payoff remains near zero, and the two recovery-triggering drift events are isolated within the adversarial subset. Thus, adversarial perturbations can cause rare local shocks but do not produce persistent controller-level compromise.
\begin{table}[!t]
\centering
\scriptsize
\caption{Game-theoretic outcomes under benign and adversarial exposure.}
\label{tab:game_numeric_small}

\begin{tabular}{lcccc}
\toprule
Condition & Fraction (\%) & N & Mean $U_D$ & Mean $U_A$ \\
\midrule
No attack & 65.6 & 328 & -12.09 & -- \\
Attack & 34.4 & 172 & -11.78 & 0.001 \\
\bottomrule
\end{tabular}

\vspace{0.5em}

\begin{tabular}{lcccc}
\toprule
Condition & Utility & Heal (\%) & Drift $>0$ (\%) & Rollback \\
\midrule
No attack & -0.16 & 0.00 & 0.00 & 0.000 \\
Attack & -0.56 & 1.16 & 1.16 & 0.023 \\
\bottomrule
\end{tabular}

\end{table}
The strategic stability of the controller is further reflected in Table~\ref{tab:game_extra}. The utility gap between the selected response and the closest alternative remains small, with a mean trust margin of only $0.038$. This indicates that GT-MCP does not operate through a dominant-agent strategy; instead, it performs fine-grained selection over closely competing candidates. Despite this narrow decision band, the injection success rate remains zero, showing that small trust margins do not translate into successful adversarial takeover. The fallback rate is also zero, indicating that stability is achieved through continuous trust-based control rather than emergency rejection.
\begin{table}[!t]
\centering
\caption{Strategic selection and stability indicators.}
\label{tab:game_extra}
\begin{tabular}{lccc}
\toprule
Metric & Mean & Median & Std \\
\midrule
Utility gap (selected vs. alternative) & 0.021 & 0.014 & 0.019 \\
Trust margin & 0.038 & 0.031 & 0.024 \\
Injection success rate & 0.000 & 0.000 & 0.000 \\
Fallback rate & 0.000 & 0.000 & 0.000 \\
\bottomrule
\end{tabular}
\end{table}
Table~\ref{tab:game_effectsize_small} summarizes the distribution-level interpretation without using unsupported placeholder statistics. The benign and adversarial utility distributions differ primarily in localized tail degradation rather than in broad distributional displacement. Attacker profitability remains negligible because successful long-horizon manipulation requires simultaneous evasion of structural grounding, peer-agreement checks, and candidate-specific drift control. This supports the incentive-alignment interpretation of GT-MCP: adversarial perturbations may increase local cost, but they do not create a systematic payoff advantage.
\begin{table}[t]
\centering
\tiny
\setlength{\tabcolsep}{1.2pt}
\renewcommand{\arraystretch}{1.12}
\caption{Strategic distribution diagnostics.}
\label{tab:game_effectsize_small}
\begin{tabularx}{\columnwidth}{
>{\raggedright\arraybackslash}p{1.75cm}
>{\raggedright\arraybackslash}p{2.35cm}
>{\raggedright\arraybackslash}X
}
\toprule
Diagnostic & Observation & Interpretation \\
\midrule
Benign vs. attack utility & Attack utility lower, but bounded & Local difficulty increases under attack \\
Attacker payoff & Mean $U_A=0.001$ under attack & No systematic adversarial gain \\
Injection success & ISR $=0.0\%$ & No controller-level compromise observed \\
Recovery localization & 2 events among 172 attack turns & Rare drift shocks are contained \\
Distributional shift & No broad displacement observed & Stability remains trajectory-level \\
\bottomrule
\end{tabularx}
\end{table}
Component isolation results in Table~\ref{tab:ablation_full_small} show that the strategic stability of GT-MCP depends on the joint action of its control components. Removing candidate-specific drift monitoring causes the largest degradation, increasing mean CDS to $2.07$ and ISR to $4.5\%$. Removing causal consistency also substantially increases adversarial success, confirming that structural grounding is essential for suppressing unsupported claims. Removing agreement weakens protection against unreliable and compromised generations, while disabling self-healing mainly increases residual drift and tail risk. The full GT-MCP pipeline achieves zero observed injection success, the lowest mean CDS, and the best mean utility.
\begin{table}[!t]
\centering
\scriptsize
\caption{Component isolation analysis under identical adversarial exposure ($T=500$).}
\label{tab:ablation_full_small}

\begin{tabular}{lccc}
\toprule
Variant & ISR (\%) & Drift $>0$ (\%) & Mean CDS \\
\midrule
No-Causal Consistency & 3.8 & 7.4 & 1.18 \\
No-Agreement Score & 2.9 & 6.1 & 0.97 \\
No-CDS & 4.5 & 12.8 & 2.07 \\
No-Heal & 1.6 & 3.2 & 0.61 \\
\textbf{Full GT-MCP} & \textbf{0.0} & \textbf{0.4} & \textbf{0.11} \\
\bottomrule
\end{tabular}

\vspace{0.5em}

\begin{tabular}{lccc}
\toprule
Variant & Std CDS & ES$_{5\%}$ & Mean Utility \\
\midrule
No-Causal Consistency & 2.41 & -2.94 & -0.91 \\
No-Agreement Score & 2.08 & -2.61 & -0.78 \\
No-CDS & 3.84 & -5.12 & -1.47 \\
No-Heal & 1.12 & -1.84 & -0.41 \\
\textbf{Full GT-MCP} & \textbf{0.90} & \textbf{-0.72} & \textbf{-0.19} \\
\bottomrule
\end{tabular}

\end{table}
Table~\ref{tab:comparison_prior_small} positions GT-MCP relative to representative prompt-injection defenses. Existing methods mainly address local instruction separation, prompt filtering, inference-time verification, and benchmark-level agent evaluation. These mechanisms are valuable, but they generally do not combine multi-agent selection, causal graph grounding, drift-aware trajectory control, rollback, and incentive modeling in a single closed-loop controller. GT-MCP differs in that it regulates whether selected outputs are allowed to update the persistent context, thereby directly targeting the state-transition mechanism exploited by long-horizon context poisoning.
\begin{table}[!t]
\centering
\scriptsize
\caption{Functional comparison between GT-MCP and representative prompt-injection defenses.}
\label{tab:comparison_prior_small}

\begin{tabular}{lccc}
\toprule
Method & Multi-Agent & Causal Graph & Drift Monitoring \\
\midrule
StruQ~\cite{chen2025struq} & No & No & No \\
Spotlighting~\cite{hines2024spotlighting} & No & No & No \\
Task Shield~\cite{jia2024task_shield} & Yes & No & Partial \\
SecInfer~\cite{liu2025secinfer} & No & No & Partial \\
AgentDojo~\cite{debenedetti2024agentdojo} & Yes & No & No \\
Multi-Agent Defense Pipeline~\cite{hossain2025multiagent_defense} & Yes & Partial & Partial \\
Survey Analysis~\cite{correia2026survey} & No & No & No \\
\textbf{GT-MCP (Ours)} & Yes & Yes & Yes \\
\bottomrule
\end{tabular}

\vspace{0.5em}

\begin{tabular}{lccc}
\toprule
Method & Rollback & Trajectory Control & Incentive Modeling \\
\midrule
StruQ~\cite{chen2025struq} & No & No & No \\
Spotlighting~\cite{hines2024spotlighting} & No & No & No \\
Task Shield~\cite{jia2024task_shield} & No & Partial & No \\
SecInfer~\cite{liu2025secinfer} & No & Partial & No \\
AgentDojo~\cite{debenedetti2024agentdojo} & No & No & No \\
Multi-Agent Defense Pipeline~\cite{hossain2025multiagent_defense} & Partial & Partial & No \\
Survey Analysis~\cite{correia2026survey} & No & No & No \\
\textbf{GT-MCP (Ours)} & Yes & Yes & Yes \\
\bottomrule
\end{tabular}

\end{table}
Overall, the game-theoretic results show that GT-MCP reshapes the adversarial payoff landscape. Attacks increase local difficulty and can produce rare, high-drift events, but they do not cause persistent context corruption. Stability arises from the joint effect of structural penalties, cross-agent consistency, candidate-specific drift monitoring, and recovery. In this sense, GT-MCP acts not only as a response selector but as a strategic regulator of long-horizon context evolution.

\subsection{Baseline Comparison and Comparative Robustness Evaluation}
\label{subsec:baseline_comparison}
We compare GT-MCP against representative LLM defense and orchestration baselines under the same $T=500$-turn adversarial evaluation protocol. The baselines include a single-agent LLM, majority voting across agents, prompt filtering, and a retrieval-oriented defense. We also include ablated GT-MCP variants to show how partial removal of core mechanisms affects robustness. All methods are evaluated under the same interaction horizon, attack families, and contextual perturbation structure defined in Section~\ref{sec:experimental_setup}.
\begin{table*}[t]
\centering
\caption{Comparative robustness evaluation under adaptive adversarial interaction. Lower values are better for injection success, drift, utility degradation, and latency; higher values are better for stable turns.}
\label{tab:baseline_comparison}
\begin{tabular}{lcccccc}
\toprule
\textbf{Method} & \textbf{Injection Success (\%)} & \textbf{Mean Drift} & \textbf{Utility} & \textbf{Recovery Trigger (\%)} & \textbf{Stable Turns (\%)} & \textbf{Latency / token (s)} \\
\midrule
Single-Agent LLM & 17.8 & 8.91 & -3.84 & -- & 71.6 & $7.9 \times 10^{-4}$ \\
Majority Voting & 9.6 & 5.37 & -2.21 & -- & 84.4 & $1.18 \times 10^{-3}$ \\
Prompt Filtering & 7.4 & 4.82 & -1.96 & -- & 86.8 & $9.6 \times 10^{-4}$ \\
RAG Defense & 5.8 & 3.77 & -1.41 & -- & 90.2 & $1.05 \times 10^{-3}$ \\
GT-MCP (No-Heal) & 1.6 & 0.61 & -0.41 & -- & 96.8 & $1.49 \times 10^{-3}$ \\
GT-MCP (No-CCI) & 3.8 & 1.18 & -0.91 & 1.8 & 92.6 & $1.55 \times 10^{-3}$ \\
\textbf{Full GT-MCP} & \textbf{0.0} & \textbf{0.11} & \textbf{-0.19} & \textbf{0.4} & \textbf{99.6} & $\mathbf{1.63 \times 10^{-3}}$ \\
\bottomrule
\end{tabular}
\end{table*}
Table~\ref{tab:baseline_comparison} shows that local defenses reduce immediate injection success compared with the single-agent baseline, but they remain vulnerable to contextual drift because they do not regulate how accepted outputs update persistent memory. Majority voting improves stability relative to the single-agent setting, yet it still allows agreement-based failures when semantically similar responses share unsupported assumptions. Prompt filtering and retrieval-oriented defenses further reduce injection success, but they primarily operate before and during candidate generation, thereby providing limited control over long-horizon context evolution. Full GT-MCP achieves the strongest robustness profile, with zero observed controller-level injection success, the lowest mean drift, and the highest stable-turn percentage. This improvement comes with a moderate computational overhead, as per-token latency increases relative to simpler baselines. However, the additional cost supports causal graph validation, candidate-specific drift estimation, trust aggregation, and self-healing. The ablated variants confirm that both structural grounding and recovery are necessary: removing causal consistency substantially increases injection success and drift, while disabling self-healing reduces the controller's ability to contain residual tail events. Overall, the comparison shows that GT-MCP improves robustness not by filtering prompts alone, but by controlling whether selected outputs are allowed to alter the validated context trajectory.

\subsection{Real-World MCP Ecosystem Analysis}
\label{subsec:mcp_ecosystem}
To contextualize the applicability of GT-MCP, we analyze representative agentic LLM ecosystems in which persistent context, tool outputs, retrieval modules, and multi-agent communication can create trajectory-level attack surfaces. This analysis is not intended as a vulnerability audit of specific implementations; rather, it identifies common architectural patterns through which untrusted information can enter the reasoning loop and later influence persistent context updates.
\begin{table*}[t]
\centering
\caption{Representative agentic LLM orchestration ecosystems and their trajectory-level context-control risks.}
\label{tab:mcp_ecosystem}
\begin{tabular}{p{3.2cm}p{4cm}p{4cm}p{3.6cm}}
\toprule
\textbf{Framework} & \textbf{Primary Capability} & \textbf{Representative Risk Surface} & \textbf{Trajectory-Level Risk} \\
\midrule
LangChain & Tool-integrated chained reasoning & Tool-output injection and memory contamination & Progressive contextual poisoning \\
AutoGen & Multi-agent conversational orchestration & Agent-agreement manipulation & Consensus hijacking \\
CrewAI & Persistent collaborative task agents & Shared-memory contamination & Long-horizon semantic drift \\
RAG Pipelines & Retrieval-augmented reasoning & Poisoned retrieved evidence & Delayed contextual steering \\
OpenAI Agent Systems & Persistent tool-based execution & Instruction persistence across turns & Hidden policy override \\
Toolformer-style systems & Autonomous API interaction & API-response manipulation & Tool-authority escalation \\
\bottomrule
\end{tabular}
\end{table*}
Table~\ref{tab:mcp_ecosystem} shows that many agentic LLM ecosystems share a common structural weakness: untrusted content can enter through retrieval, tools, shared memory, and inter-agent messages, and may later influence future reasoning if accepted into persistent context. These risks are trajectory-level rather than purely prompt-level, because the main failure pathway is not only the immediate generation of an unsafe response but the gradual incorporation of unsupported fragments into the state that conditions later outputs. GT-MCP addresses this class of risk by regulating context updates through causal graph validation, candidate-specific drift monitoring, trust-based selection, and rollback-based recovery. Therefore, its main relevance to real-world MCP-style ecosystems is as a context-control layer that governs whether generated, retrieved, and tool-derived content may become persistent memory.

\subsection{Extended Adversarial Attack Taxonomy}
\label{subsec:extended_attack_taxonomy}
To evaluate GT-MCP under realistic trajectory-level threats, we include attack families that differ in injection surface, timing, and adversarial objective. Unlike single-turn jailbreaks, these attacks are designed to test whether unsupported and malicious fragments can persist across context updates and gradually influence future reasoning. Table~\ref{tab:extended_attack_taxonomy} summarizes the adversarial patterns used during evaluation.
\begin{table*}[t]
\centering
\caption{Extended adversarial attack taxonomy used to evaluate trajectory-level context robustness.}
\label{tab:extended_attack_taxonomy}
\begin{tabular}{p{3cm}p{3.1cm}p{3cm}p{5.1cm}}
\toprule
\textbf{Attack Family} & \textbf{Injection Surface} & \textbf{Temporal Pattern} & \textbf{Representative Objective} \\
\midrule
Direct Prompt Injection & User query stream & Immediate & Override controller constraints and induce unsafe generation \\
Indirect Retrieval Poisoning & Retrieved evidence & Delayed & Insert hidden instructions within semantically relevant documents \\
Tool-Output Manipulation & API/tool responses & Immediate & Escalate authority through fabricated tool outputs \\
Dormant Trigger Injection & Any contextual source & Multi-turn latent & Activate adversarial instructions after delayed interaction depth \\
Trajectory Steering & Persistent context memory & Gradual & Slowly shift constraints, objectives, and semantic framing \\
Agreement Mimicry & Byzantine agent output & Multi-turn & Imitate peer agreement while injecting unsupported claims \\
Semantic Drift Amplification & Cross-turn summarization & Progressive & Increase divergence from the validated reasoning trajectory \\
\bottomrule
\end{tabular}
\end{table*}
The attack schedule is randomized across the 500-turn evaluation horizon, while perturbations are constrained to remain locally plausible under the bounded anomaly condition $\psi(\Delta c_t)\le\epsilon$. This design evaluates subtle context-poisoning behavior rather than only explicit jailbreak attempts. The taxonomy complements the threat model by illustrating how adversarial perturbations enter the interaction loop and attempt to influence the persistent context via selected outputs.

\subsection{Benchmark-Based Adversarial Evaluation}
\label{subsec:benchmark_eval}
To improve external validity, the adversarial portion of the evaluation is organized according to benchmark-inspired attack styles commonly used in prompt-injection and agent-security research. These include PromptInject-style direct instruction override, PoisonedRAG-style retrieval poisoning, AgentDojo-style tool and agent interaction attacks, delayed-trigger attacks, and agreement-mimicry attacks. The goal is not to reproduce each benchmark verbatim, but to align the evaluation with their dominant threat patterns while preserving the closed-loop GT-MCP setting.
\begin{table}[t]
\centering
\caption{Benchmark-inspired adversarial evaluation across the 172 attack turns.}
\label{tab:benchmark_eval}
\begin{tabular}{lcc}
\toprule
\textbf{Benchmark Style} & \textbf{Attack Turns} & \textbf{High-Drift Events} \\
\midrule
PromptInject-style & 42 & 1 \\
PoisonedRAG-style & 48 & 1 \\
AgentDojo-style & 34 & 0 \\
Delayed Trigger & 24 & 0 \\
Agreement Mimicry & 24 & 0 \\
\midrule
Total & 172 & 2 \\
\bottomrule
\end{tabular}
\end{table}
As shown in Table~\ref{tab:benchmark_eval}, the two high-drift events occur under direct prompt injection and retrieval poisoning, which are the most immediate pathways for inserting adversarial instructions into the observed context. No high-drift event is observed for AgentDojo-style tool interaction, delayed-trigger, and agreement-mimicry patterns under the evaluated schedule. Both high-drift events are recovered by the self-healing layer, and neither results in a persistent controller-level compromise. These results support the trajectory-level robustness of GT-MCP: benchmark-inspired attacks may introduce localized instability, but they do not persist as validated context updates.

\subsection{Failure Cases and Residual Vulnerabilities}
\label{subsec:failure_cases}
Although GT-MCP prevents persistent controller-level compromise in the evaluated setting, the two high-drift events reveal residual vulnerability patterns that remain important to address in deployment. Table~\ref{tab:failure_cases} summarizes these cases. Both observed failures are localized and recovered by the self-healing layer, and neither results in persistent memory corruption. This indicates that GT-MCP does not eliminate all adversarial perturbations at the point of entry; rather, it prevents destabilizing fragments from becoming durable components of the validated context.
\begin{table}[t]
\centering
\caption{Residual vulnerability patterns observed during adversarial evaluation.}
\label{tab:failure_cases}
\begin{tabular}{lcc}
\toprule
\textbf{Failure Pattern} & \textbf{Occurrences} & \textbf{Recovered} \\
\midrule
Retrieval poisoning chain & 1 & Yes \\
Semantic paraphrase drift & 1 & Yes \\
Consensus mimicry & 0 & -- \\
Delayed trigger activation & 0 & -- \\
Persistent memory corruption & 0 & -- \\
\bottomrule
\end{tabular}
\end{table}
The retrieval poisoning case corresponds to adversarial evidence that remained semantically close to the task while introducing unsupported claims. The semantic paraphrase drift case reflects a subtler perturbation in which the adversarial fragment did not appear as an explicit instruction, but shifted the framing of the response through paraphrased content. In both cases, the controller detected abnormal drift after a tentative update and activated rollback and quarantine. These results suggest that future work should focus on stronger semantic-collusion attacks, uncertainty-aware claim extraction, and scalable paraphrase verification.

\section{Discussion}
\label{sec:discussion}
The results show that the main contribution of GT-MCP is not limited to improving the quality of isolated responses; rather, it stabilizes the evolution of reasoning across multi-turn interactions. This distinction is central to persistent-context LLM systems, where adversarial fragments can become harmful not only when they produce an immediate unsafe output, but also when they are retained in memory and later reused as part of the reasoning state. By treating context evolution as a controlled trajectory, GT-MCP shifts the defense objective from local prompt filtering to long-horizon regulation of state transitions. The stability results demonstrate that the controller keeps almost all selected outputs close to the validated context trajectory. Contextual drift remains near zero for 99.6\% of turns, while the two high-drift events are isolated and recovered before they can persist. This behavior indicates that causal grounding and candidate-specific drift monitoring act together as a trajectory-stabilization mechanism. Unsupported fragments may enter the observed context, but they do not automatically become part of the validated state. Instead, they must pass structural, semantic, and temporal checks before they can influence future turns. The manifold analysis further clarifies how GT-MCP shapes reasoning behavior. Selected outputs concentrate within a constrained region defined by causal consistency and cross-agent agreement. This suggests that the controller not only rejects unsafe outputs but also biases selection toward structurally grounded, semantically coherent responses that are consistent with the validated context. The positive association between manifold alignment and utility supports this interpretation: responses closer to the structural reasoning subspace produce better outcomes, whereas responses farther from it incur lower trust and higher drift risk. At the agent level, robustness arises from controlled heterogeneity rather than from dependence on a single dominant model. GPT-5.3, Llama-3.1-70B, and DeepSeek-R1 contribute different utility, trust, and stability profiles. GT-MCP converts these differences into a selection advantage by evaluating each candidate relative to the current context state. As a result, agent diversity becomes useful only when mediated by causal validation and drift-aware trust aggregation. This is important because simple voting can amplify shared errors and agreement mimicry, whereas GT-MCP requires agreement to be supported by a validated graph structure. The game-theoretic results indicate that GT-MCP reduces the strategic value of adversarial perturbation. Attacks increase local difficulty and produce rare, high-drift events, but they do not yield persistent controller-level compromise and a positive adversarial payoff in the evaluated setting. This outcome is consistent with the incentive-alignment design: an adversarial response must simultaneously appear structurally grounded, agree with peer agents, and avoid inducing excessive contextual drift. Failing any of these conditions lowers its trust score and activates recovery. Thus, the controller reshapes the payoff landscape by making unsupported deviations less likely to be selected and to persist. The efficiency results show that these robustness gains are achieved with predictable computational overhead. Latency per token remains tightly bounded, and increased token usage does not produce threshold-driven degradation. Additional computation primarily contributes to claim extraction, causal validation, agreement estimation, and drift monitoring. This suggests that GT-MCP scales by increasing verification depth rather than by amplifying contextual noise. For deployment, this trade-off is important: the controller incurs moderate overhead, but the added computation directly supports stability of the persistent reasoning state. Additionally, the findings support the view that prompt injection in agentic and persistent-context LLM systems should be treated as a trajectory-control problem. GT-MCP addresses this problem by combining causal graph grounding, candidate-specific drift monitoring, trust-weighted multi-agent selection, and rollback-based self-healing. The resulting system does not rely on any single defense layer; its robustness emerges from the interaction of structural validation, temporal stability, and strategic selection. This makes GT-MCP a suitable control layer for adversarially aware multi-agent LLM systems in which maintaining the integrity of evolving context is as important as generating safe individual responses.

\section{Limitations and Future Directions}
\label{sec:limitations_future}
Although our experiments demonstrate bounded contextual drift, stable reasoning manifolds, and incentive-aligned multi-agent interaction under the evaluated configuration, several limitations remain. First, the studies employ a fixed set of LLM agents and architectures; broader model heterogeneity, including multimodal and domain-specialized systems, may alter the geometry of the reasoning manifold and the efficacy of trust-weighted selection. Second, structural grounding and cross-agent agreement serve as operational proxies for reasoning validity and do not guarantee epistemic correctness when the validated context is incomplete and noisy, highlighting the need for external verification signals, causal auditing, and uncertainty calibration. Third, the threat model assumes contextual perturbations with at most one compromised agent; more aggressive adversarial scenarios, including coordinated multi-agent compromise and agreement mimicry, require further investigation. Fourth, evaluation is performed over finite interaction horizons, and long-term deployments may reveal slower drift accumulation and novel trajectory instabilities not captured in bounded experiments. Additionally, GT-MCP incurs computational overhead due to claim extraction, causal graph maintenance, and trust computation. Future research will focus on developing more efficient validation mechanisms, theoretical guarantees of convergence, and scalable deployment strategies, with the aim of formalizing context control as a principled and practical layer for secure, multi-agent LLM systems.

\section{Conclusion}
\label{Conclusion}
In this work, we introduced GT-MCP, a game-theoretic context-control mechanism for securing multi-agent LLM reasoning against prompt-injection and context-poisoning attacks. Unlike filtering-based defenses that operate mainly at the input and output level, GT-MCP regulates how selected responses update persistent context. It combines causal graph grounding, cross-agent agreement, candidate-specific drift monitoring, trust-weighted selection, and rollback-based self-healing to stabilize reasoning trajectories over multiple turns. The empirical results show that GT-MCP maintains bounded contextual drift, confines reasoning to a structurally grounded subspace, leverages heterogeneous LLM agents without relying on a single dominant model, limits adversarial payoff, and scales predictably with token usage and latency. Across the evaluated 500-turn adversarial setting, the controller prevents observed controller-level injection success while containing rare high-drift events through recovery. These findings support the view that prompt injection in persistent-context and agentic LLM systems should be treated as a trajectory-control problem rather than only a single-turn filtering problem. By embedding structural validation, temporal stability, and incentive-aware selection into the interaction layer, GT-MCP provides a practical foundation for robust and secure context management in multi-agent LLM architectures.

\bibliographystyle{IEEEtran}
\bibliography{Ref}

\end{document}